\documentclass[a4wide,10pt,onecolumn]{article}
\usepackage{amsmath,amsfonts,amssymb}
\usepackage{pstricks}
\usepackage{graphicx}
\usepackage{fullpage}
\usepackage{authblk}
\usepackage[utf8]{inputenc}

\begin{document}
\title{Numerical Results for the System Noise Temperature of an Aperture Array Tile and Comparison with Measurements \vspace{0.5cm}}
\date{October 15, 2011}
\author[1]{M.V.\ Ivashina \thanks{ivashina@chalmers.se}}
\author[2]{E.E.M.\ Woestenburg \thanks{woestenburg@astron.nl}}
\author[2]{L.\ Bakker \thanks{bakker@astron.nl}}
\author[2]{R.H.\ Witvers \thanks{witvers@astron.nl} \vspace{0.5cm}}
\affil[1]{Department of Earth and Space Sciences, Chalmers University of Technology, Sweden}
\affil[2]{The Netherlands Institue for Radio Astronomy (ASTRON)}
\maketitle

\section{Introduction}

This work has been supported in part by  the Netherlands Organization for Scientific Research (APERTIF project funded by NWOGroot),
the Swedish Agency for Innovation Systems VINNOVA and Chalmers University of Technology (VINNMER - Marie Curie Actions international qualification fellowship).
Some results of this work have been recently published in~\cite{WoestenburgMW:2011,Woestenburg:2011}.

This report is a shortened version of the original ASTRON report (RP324, 2011) which is publically availible upon request.
The purpose of this report is to document the noise performance of the
Apertif array receiver which has been used to measure its performance
as an aperture array and to characterize the recently developed noise 
measurement facility THACO. The receiver system is the second prototype 
of APERTIF (DIGESTIF2)~\cite{Verheijen:2008,CappellenAPS:2009} that includes the array
antenna of 144 dual-polarized TSA elements, 144 Low Noise Amplifiers 
(LNAs) ($T_\text{min}=$35-40K) and the data recording/storing facilities of
the initial test station that allow for off-line digital beamforming.

The primary goal of this study is to compare the measured receiver
noise temperatures with the simulated values for several practical
beamformers, and to predict the associated receiver noise coupling
contribution, antenna thermal noise and ground noise pick-up (due to
the back radiation). The measurements were performed over a wide
frequency band and scan range. In the course of the measurements,
the 4, 16, 25 and 49 active antenna elements were used in
beamforming, while the remaining elements were connected to the LNAs
with the outputs terminated in matched loads. The experimental results
were obtained when the receiver with 4- and 16-element analog 
beamformers was placed inside THACO (a metal shielding cabin which 
was designed to mitigate the noise contributions due to the ground and 
trees) and in the open environment at Westerbork near the Westerbork
Synthesis Radio Telescope (WSRT), using digital beamforming. The details
of these measurements can be found in~\cite{Woestenburg:2011}.

Since, the digital data recording-storing facilities are available
for the DIGESTIF array receiver, one can also evaluate the system
noise temperature for beamforming scenarios that are based on
standard signal processing
algorithms~\cite{VanTrees:2002}--\cite{IvashinaJAP:2010} and the
noise correlation coefficients between the array channels. In this
report, we therefore show the range of the realized system noise
temperatures when an embedded element only and all array elements
are used in beamforming. These results demonstrate the \textit{pros}
and \textit{cons} of the considered low-gain antenna and high gain
digital array receivers for the purpose of the noise temperature
characterization in an open environment and inside a shielding cabin
such as THACO.

\section{The model of the array antenna}
The array consists of $2\times 72$ aluminium Tapered Slot Antenna
(TSA) elements with a pitch of 11 cm ($0.52\lambda$ at 1420 MHz) on
a rectangular $8\times 9$ grid. Each TSA is fed by a wideband
microstrip feed which has been integrated with an LNA on a printed
circuit board. This design features a very short transmission line
between the antenna and the LNA, since the circular slotline cavity
has been moved sideways.
More details on the array design and the numerical approach used for the EM-analysis of the antenna can be found
in~\cite{ArtsEuCAP:2010} and~\cite{IvashinaJAP:2010,IvashinaICEAA:2010}, respectively. 
The simulations of the antenna have been carried out using CAESAR software that is an array system simulator, developed at ASTRON~\cite{Maaskant:2006}.
Note that these simulations did not take into account the effect of the contributions from obstacles (trees and
telescopes/buildings) near the horizon. 

\section{The noise model of the receiver system}
The noise model used in this study is based on the equivalent system
representation as described in~\cite{Ivashina:2008}. According to
this representation, the sensitivity of the array receiver can be
computed as follows:
\begin{equation}\label{Sensitivity_Sigma1}
\frac{A_{eff}}{T_{sys}}=\displaystyle\frac{A_{ph}\eta_{ap}}{T_{ext}+\displaystyle\left(\frac{1-\eta_{rad}}{\eta_{rad}}\right)T_{amb}+\left(\frac{1}{\eta_{rad}}\right)T^{LNA}_{Eq}}.
\end{equation}
where $A_\text{ph}$ and $T_\text{sys}$ are the physical antenna area
and system noise temperature, respectively. The latter consists of
three main contributions: (i) The external noise contribution - the ground noise picked up due
to antenna back radiation, which was computed from the simulated illumination pattern of the antenna array for the
specified beam former weights, (ii) the thermal antenna noise due to the losses in the
conductor and dielectric materials of TSAs and
microstrip feeds. The conductor losses are computed
through the evaluation of the antenna radiation efficiency
using the methodology detailed in~\cite{MaaskantAWPL:2009} and the dielectric
losses are computed based on the experimental
evaluation of the feed loss~\cite{Ivashina:2007} and
(iii) the noise due to LNAs which is dependent upon the noise
properties of LNAs and active reflection coefficients seen at the
ports of the antenna array. Note, that by using this definition of
$T_{sys}$, all noise temperature contributions are referenced to the
sky (in front of the antenna aperture).

It is important to note that the system noise temperature which is
calculated from the measured Y-factor is also referenced to the sky.
Therefore, to distinguish between the external noise, antenna
thermal noise and receiver contributions, one needs to know the beam
shape (to compute $T_\text{ext}$) and radiation efficiency
$\eta_\text{rad}$ of the antenna, which are in general dependent on
the beamformer weights.

\section{The model of the beamformer used to compute the optimal weights}
The beamformer model used in this study is based on the mathematical
framework which has been described in~\cite{IvashinaJAP:2010}.
Within this framework, the array receiver system is subdivided into
two blocks: (i) the front-end including the array antenna, Low Noise
Amplifiers (LNAs); and (ii) the beamformer with complex conjugated
weights $\{w_{n}^{*}\}_{n=1}^{N}$ and an ideal
(noiseless/reflectionless) power combiner realized in software.
Here, $\boldsymbol{\mathsf{w}}^{H}=[w_{1}^{*},\ldots,w_{N}^{*}]$ is
the beamformer weight vector, $H$ is the Hermitian transpose, and
the asterisk denotes the complex conjugate. Furthermore,
$\boldsymbol{\mathsf{a}}=[a_{1},\ldots,a_{N}]^{T}$ is the vector
holding the transmission-line voltage-wave amplitudes at the
beamformer input (the $N$ LNAs outputs). Hence, the fictitious
beamformer output voltage $v$ (across $Z_{0}$) can be written as
$v=\boldsymbol{\mathsf{w}}^{H}\boldsymbol{\mathsf{a}}$, and the
receiver output power as
$|v|^{2}=vv^{*}=(\boldsymbol{\mathsf{w}}^{H}\boldsymbol{\mathsf{a}})(\boldsymbol{\mathsf{w}}^{H}\boldsymbol{\mathsf{a}})^{*}=(\boldsymbol{\mathsf{w}}^{H}\boldsymbol{\mathsf{a}})(\boldsymbol{\mathsf{a}}^{T}\boldsymbol{\mathsf{w}}^{*})^{*}=\boldsymbol{\mathsf{w}}^{H}\boldsymbol{\mathsf{a}}\boldsymbol{\mathsf{a}}^{H}\boldsymbol{\mathsf{w}}$,
where the proportionality constant has been dropped as this is
customary in array signal processing and because we will consider
only ratios of powers.

Although each subsystem can be rather complex and contains multiple
internal signal/noise sources, it is characterized externally (at
its accessible ports) by a scattering matrix in conjunction with a
noise- and signal-wave correlation matrix. In this manner, the
system analysis and weight optimization becomes a purely linear
microwave circuit problem.
The sensitivity metric $A_{\text{eff}}/T_{\text{sys}}$, which is the
effective area of the antenna system divided by the system
equivalent noise temperature, can be expressed in terms of the
Signal-to-Noise Ratio (SNR) and the normalized flux density
$S_{\text{source}}$ of the source (in Jansky,
$1\,[\text{Jy}]=10^{-26}\,[\text{Wm}^{-2}\text{Hz}^{-1}]$) as
\begin{align}
\frac{A_{\text{eff}}}{T_{\text{sys}}}=\frac{2k_{\text{B}}}{S_{\text{source}}}\text{SNR},\quad\text{where}\quad\text{SNR}=\frac{\boldsymbol{\mathsf{w}}^{H}\boldsymbol{\mathsf{P}}\boldsymbol{\mathsf{w}}}{\boldsymbol{\mathsf{w}}^{H}\boldsymbol{\mathsf{C}}\boldsymbol{\mathsf{w}}},\label{SNR}
\end{align}
and where $k_{B}$ is Boltzmann's constant. The $\text{SNR}$ function
is defined as a ratio of quadratic forms where
$\boldsymbol{\mathsf{C}}$ is a Hermitian spectral noise-wave
correlation matrix holding the correlation coefficients between the
array receiver channels, i.e.,
$C_{mk}=\mathsf{E}\{c_{m}c_{k}^{*}\}=\overline{c_{m}c_{k}^{*}}$ (for
$k,m=1\ldots N$). Here, $c_{m}$ is the complex-valued voltage
amplitude of the noise wave emanating from channel $m$ (see
\cite{Wedge:1992} and references therein), which includes the
external and internal noise contributions inside the frontend block. We consider only a narrow frequency band, and
assume that the statistical noise sources are (wide-sense)
stationary random processes which exhibit ergodicity, so that the
statistical expectation can be replaced by a time average (as also
exploited in hardware correlators). $\boldsymbol{\mathsf{C}}$ is
nonzero if noise sources are present in the external environment and
inside the system, due to e.g. the ground, LNAs, and sky. For a single point source on the sky, the
signal-wave correlation matrix
$\boldsymbol{\mathsf{P}}=\boldsymbol{\mathsf{e}}\boldsymbol{\mathsf{e}}^{H}$
is a one-rank positive semidefinite matrix. The vector
$\boldsymbol{\mathsf{e}}=[e_1, e_2,.., e_N]^T$ holds the signal-wave
amplitudes at the receiver outputs and arises due to an externally
applied plane electromagnetic wave $\boldsymbol{E}_\text{i}$.
\subsection{Maximum Sensitivity}
Maximizing Eq.~\eqref{SNR} amounts to solving the largest root of the
determinantal equation~\cite{VanTrees:2002}:
$\text{det}\left(\boldsymbol{\mathsf{P}}-\text{SNR}\,\boldsymbol{\mathsf{C}}\right)=0$
(\emph{cf}.~\cite{Cheng:1967}). Next, the optimum beamfomer weight
vector $\boldsymbol{\mathsf{w}}_\text{MaxSNR}$ is found through
solving the corresponding generalized eigenvalue equation
$\boldsymbol{\mathsf{P}}\,\boldsymbol{\mathsf{w}}_\text{MaxSNR}=\text{SNR}\,\boldsymbol{\mathsf{C}}\,\boldsymbol{\mathsf{w}}_\text{MaxSNR}$
for the largest eigenvalue (SNR) as determined in the previous step.
The well-known closed-form solution for the point source case, where
$\boldsymbol{\mathsf{P}}$ is of rank 1, is given
by~\cite{Cheng:1967}
\begin{align}
\boldsymbol{\mathsf{w}}_{\text{MaxSNR}}=\boldsymbol{\mathsf{C}}^{-1}\boldsymbol{\mathsf{e}},\quad\text{with}\quad\text{SNR}=\boldsymbol{\mathsf{e}}^{H}\boldsymbol{\mathsf{w}}_{\text{MaxSNR}}\label{Weigths_MaxSNR}
\end{align}
where the eigenvector $\boldsymbol{\mathsf{e}}$ corresponds to the
largest eigenvalue of $\boldsymbol{\mathsf{P}}$.
\subsection{Maximum output power or the Conjugate Field Matching (CFM) condition}
When $\boldsymbol{\mathsf{C}}$ equals the identity matrix
$\boldsymbol{\mathsf{I}}$ (thus equal and uncorrelated output noise
powers), the receiver output noise power
$\boldsymbol{\mathsf{w}}^{H}\boldsymbol{\mathsf{C}}\boldsymbol{\mathsf{w}}=\boldsymbol{\mathsf{w}}^{H}\boldsymbol{\mathsf{w}}$
becomes independent of $\boldsymbol{\mathsf{w}}$ in case its 2-norm
$\boldsymbol{\mathsf{w}}^{H}\boldsymbol{\mathsf{w}}$ is a constant
value, typically chosen to be unity. With reference
to~\eqref{Weigths_MaxSNR}, the weight vector that maximizes the
received power, and thus realizes a maximum directive gain (and
effective area) in the direction of observation, is therefore
\begin{align}
\boldsymbol{\mathsf{w}}_{\text{CFM}}=\boldsymbol{\mathsf{e}}.\label{Weigths_CFM}
\end{align}
These weights optimally satisfy the Conjugate Field Matching (CFM) 
condition~\cite{IvashinaEuCAP:2007,Ivashina:2009,Wood:1980}.
\subsection{Minimum system noise temperature}
Similarly, one can develop an expression for computing the
beamformer weights for the minimum $T_{sys}$, that is the case when
the source of interest has no contribution and thus independent on
the weights. For this case, the optimal beamformer is described as:
\begin{align}
\boldsymbol{\mathsf{w}}_{\text{MinTsys}}=\boldsymbol{\mathsf{C}}^{-1}\boldsymbol{\mathsf{e}_\text{o}}\label{Weigths_MinTsys}
\end{align}
where $\boldsymbol{\mathsf{e}_\text{o}}=\boldsymbol{\mathsf{1}}$.
\section{Numerical results for the 144-channel (full-polarization) beamformer}
\subsection{Simulation details}
The simulations were performed with the newly developed numerical
tool box for the CAESAR software ~\cite{IvashinaICEAA:2010}. This
toolbox was initially aimed at the analysis and optimization of the
PAF systems, and has been interfaced with GRASP to compute the
overall noise wave scattering matrix due to external and internal
noise sources as well as the secondary array patterns after the
scattering from the dish. For the present study, we have used the
pre-processor of this software to determine the optimal beamformer
weights (so as to account for the non-uniform noise distribution of
the environment) and to evaluate the receiver noise contributions
due to internal noise sources according to the model presented
in~\cite{Ivashina:2008} and~\cite{MaaskantAWPL:2009}.

\newpage
\subsection{The array beam noise temperature and its contributions}
\begin{figure}[!ht]
\centering
\begin{tabular}{cc}
\includegraphics[width=8 cm]{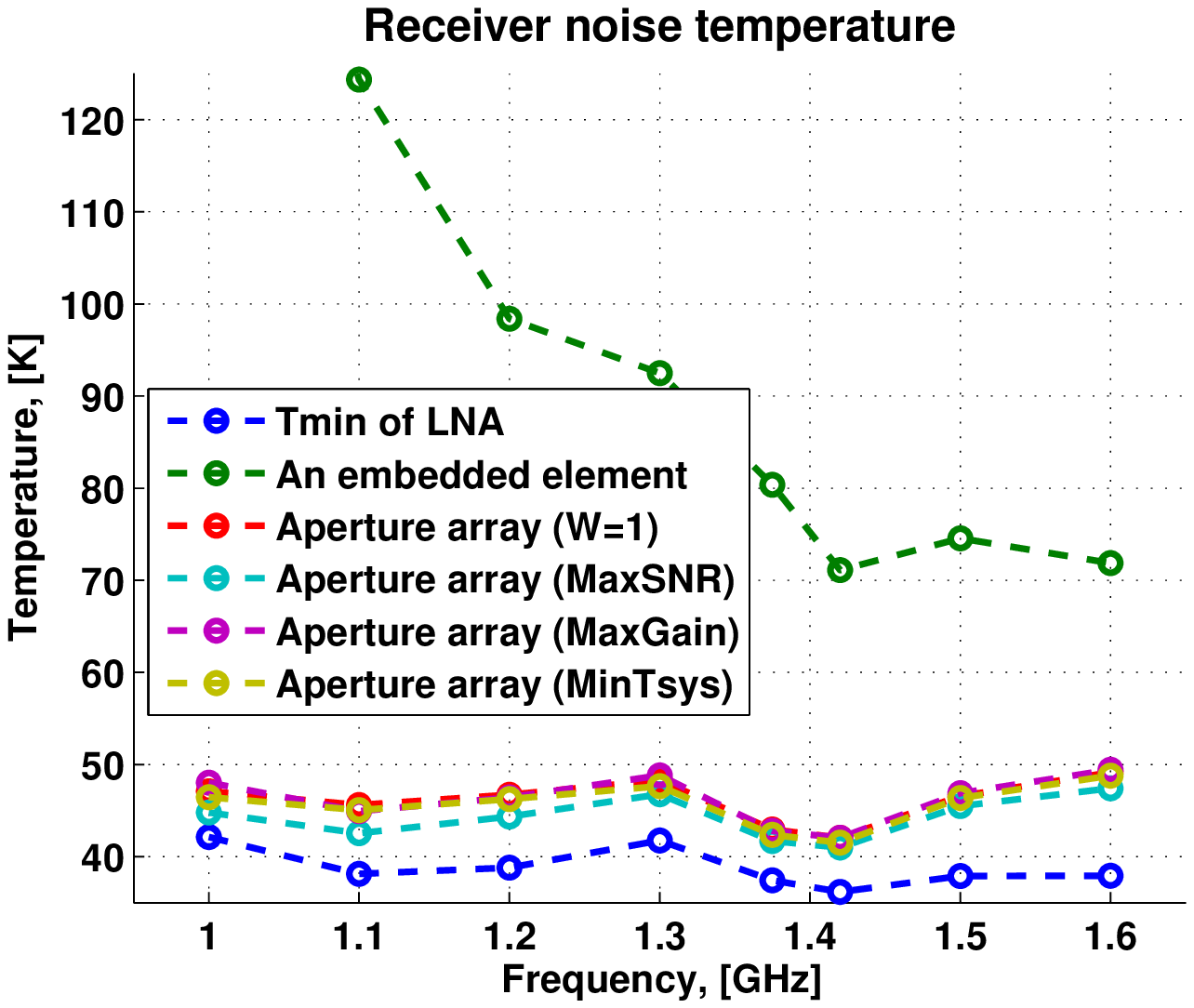} & \includegraphics[width=8cm]{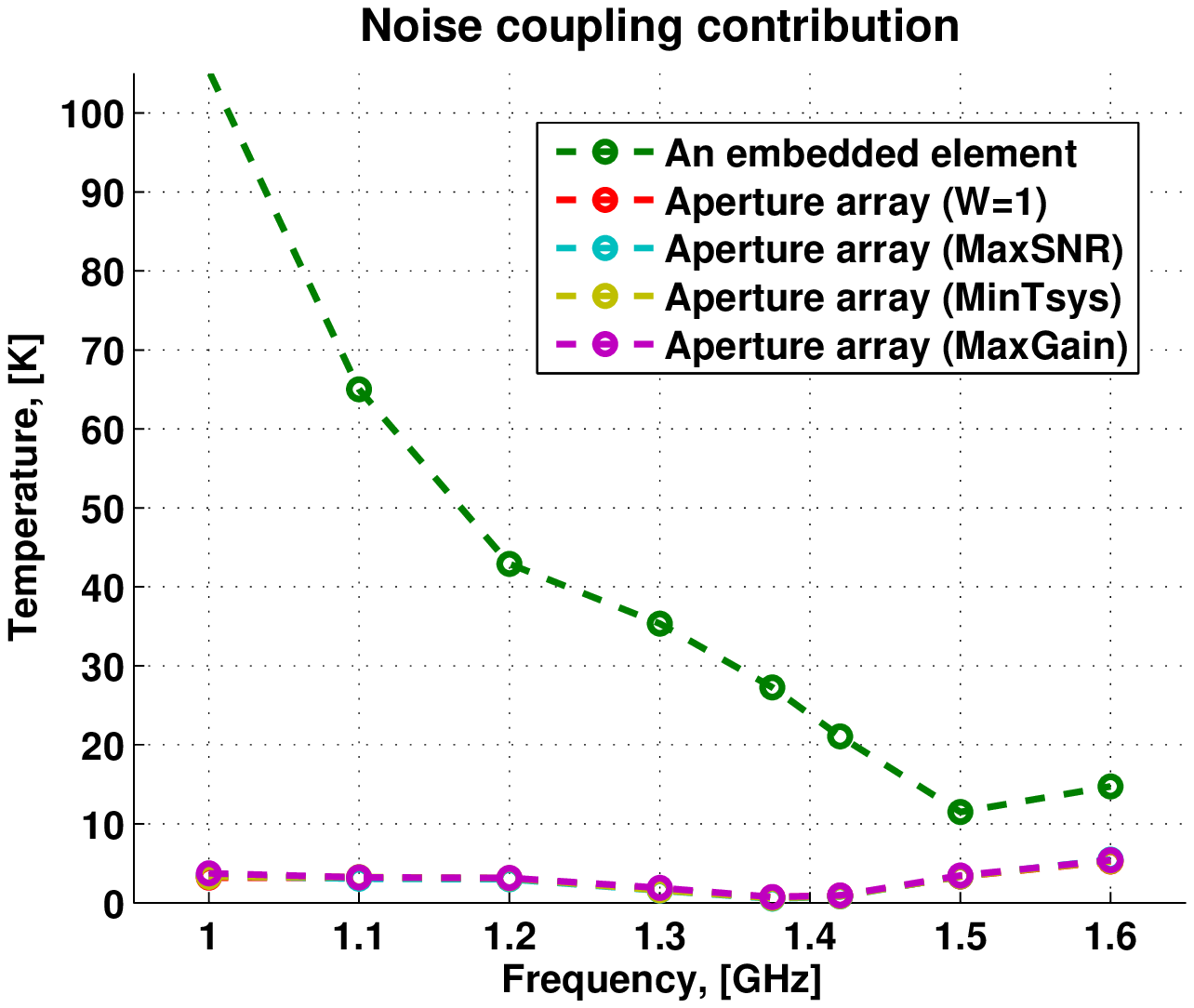}
\\
\small (a)  & \small (b)  \\\\
\includegraphics[width=8 cm]{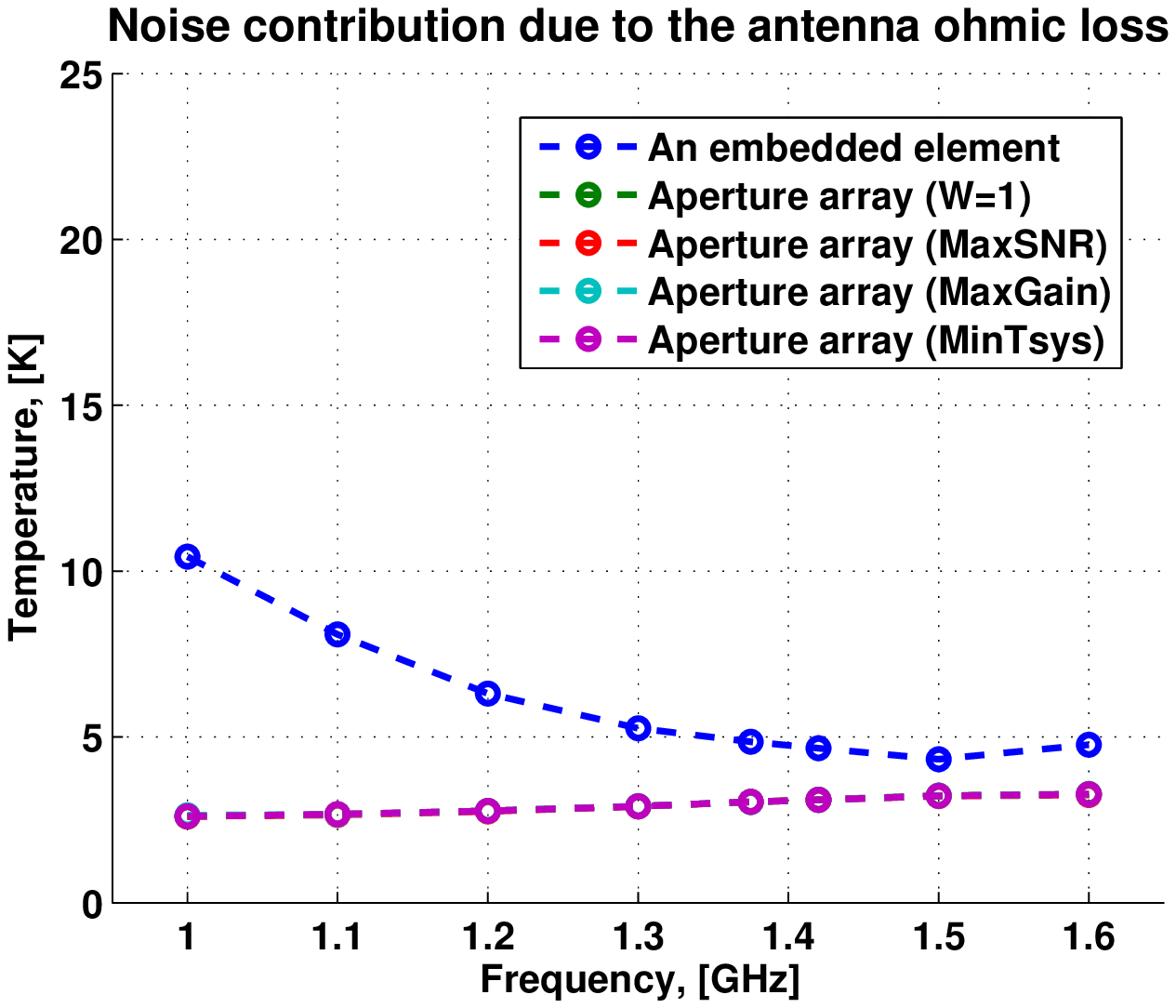} & \includegraphics[width=8cm]{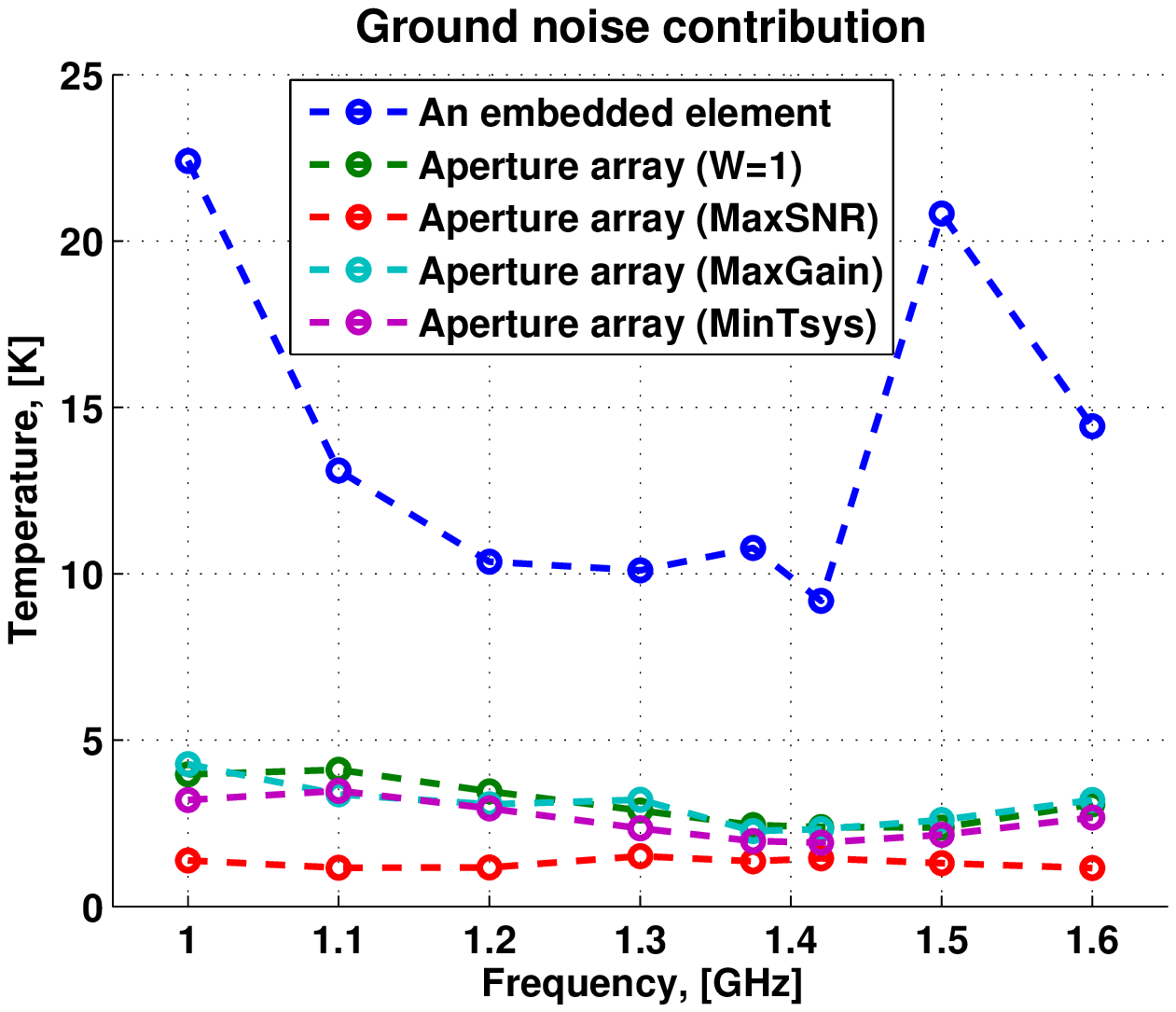}
\\
\small (c)  & \small (d)  \\\\
\end{tabular}
\caption{(a) The simulated array beam noise temperature versus
frequency and its contributions due to (b) the noise coupling
effects in the receiver, (c) the ohmic losses of the antenna and (d)
the external noise due to the ground (due to the back radiation of
the array, when $\theta\geq 90^{\text{o}}$).} \label{Tsys}
\end{figure}

\section{Results for experimental (bi-scalar) beamformers with 4, 16, 25 and 49 channels}

This chapter shows the simulated array antenna properties and compares them with
measured receiver noise temperatures versus frequency and scan angle for various
beamformer configurations. It starts with a description of the measurement setups and
method. It then presents the simulation results for the array antenna patterns and the
array beam noise temperatures and its contributions. Finally the results of
cross-comparison of the measured Trec at Westerbork and inside THACO are
presented.

\subsection{Measurement method and setups at Westerbork and inside THACO}
For the noise measurements of the APERTIF tile as an aperture array the Y-factor hot/cold 
method has been used~\cite{Woestenburg:2003}--\cite{WarnickURSI:2007}. During the
measurement the tile is placed horizontally on the ground for the measurement in Westerbork (in the open area)
or inside the big shielding facility for the Y-factor hot/cold measurements that is called THACO. Two methods were followed, one using analog
beam forming with 2x2 and 4x4 elements for the measurements inside THACO, the other
using digital beam forming for various beam configurations at the WSRT location. The output
signals from the LNAs of a 2x2 and 4x4 array in the centre of the tile inside THACO were added
with in-phase combiners, forming broadside beams. The analog output signals of the beam
formers were fed to the input of an Agilent Noise Figure Meter 8970B and the noise
temperature was determined with the Y-factor method, using the cold sky as a 'cold' load
and the roof of THACO, covered with absorbing material, as the 'hot' load.
The digital beam forming and processing of the APERTIF prototype system at the WSRT
provided a much more flexible system, with which beams could be formed with a larger number
of elements, pointing in any desired direction. For the digital processing method, a total of
49 individual antenna elements and LNAs (limited by the number of available receivers
at the time of measurements) are connected via 25 m long coaxial cables to the
back-end. The back-end electronics is located in a shielded cabin, together with the
down converter modules and digital processing hardware ~\cite{CappellenBoston:2010}.
Data are taken with the array facing the (cold) sky as 'cold' load, after which a room
temperature absorber is placed over the array for the measurement with the 'hot' load. The data processing
takes into account correlations between data from individual elements. The results are stored
in a covariance matrix as a function of frequency. Using off-line digital processing, beams
with a combination of any of the 49 active elements can be formed and beams may be
scanned in any direction by applying weights to the elements of the covariance matrix. 
In this way the equivalent beam noise temperature as a function of frequency from
1.0 to 1.8 GHz has been determined for 2x2, 4x4, 5x5 and 7x7 element arrays, looking
at broadside. Also the equivalent beam noise temperatures as a function of scan angle
for the 4x4 and 7x7 element arrays have been determined.

\subsection{Simulated antenna beam directivity and noise temperatures}
\begin{figure}[!ht]
\centering
\includegraphics[width=10 cm]{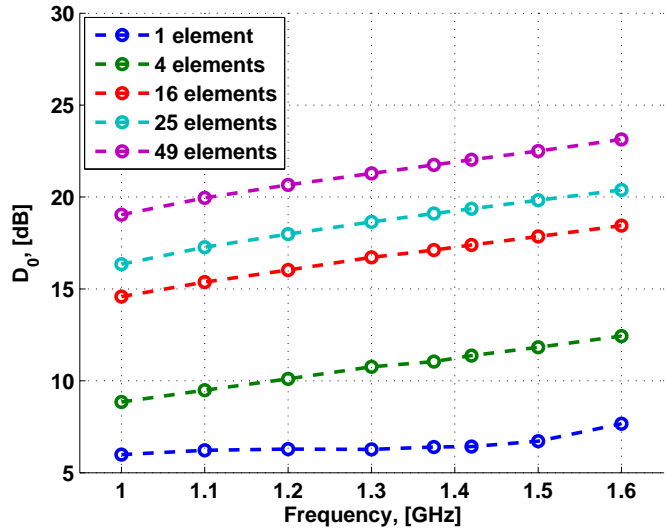}\\
 \caption{The array directivity versus frequency.} \label{ExpArray_weights}
\end{figure}

\begin{figure}[!ht]
\centering
\begin{tabular}{cc}
\includegraphics[width=8 cm]{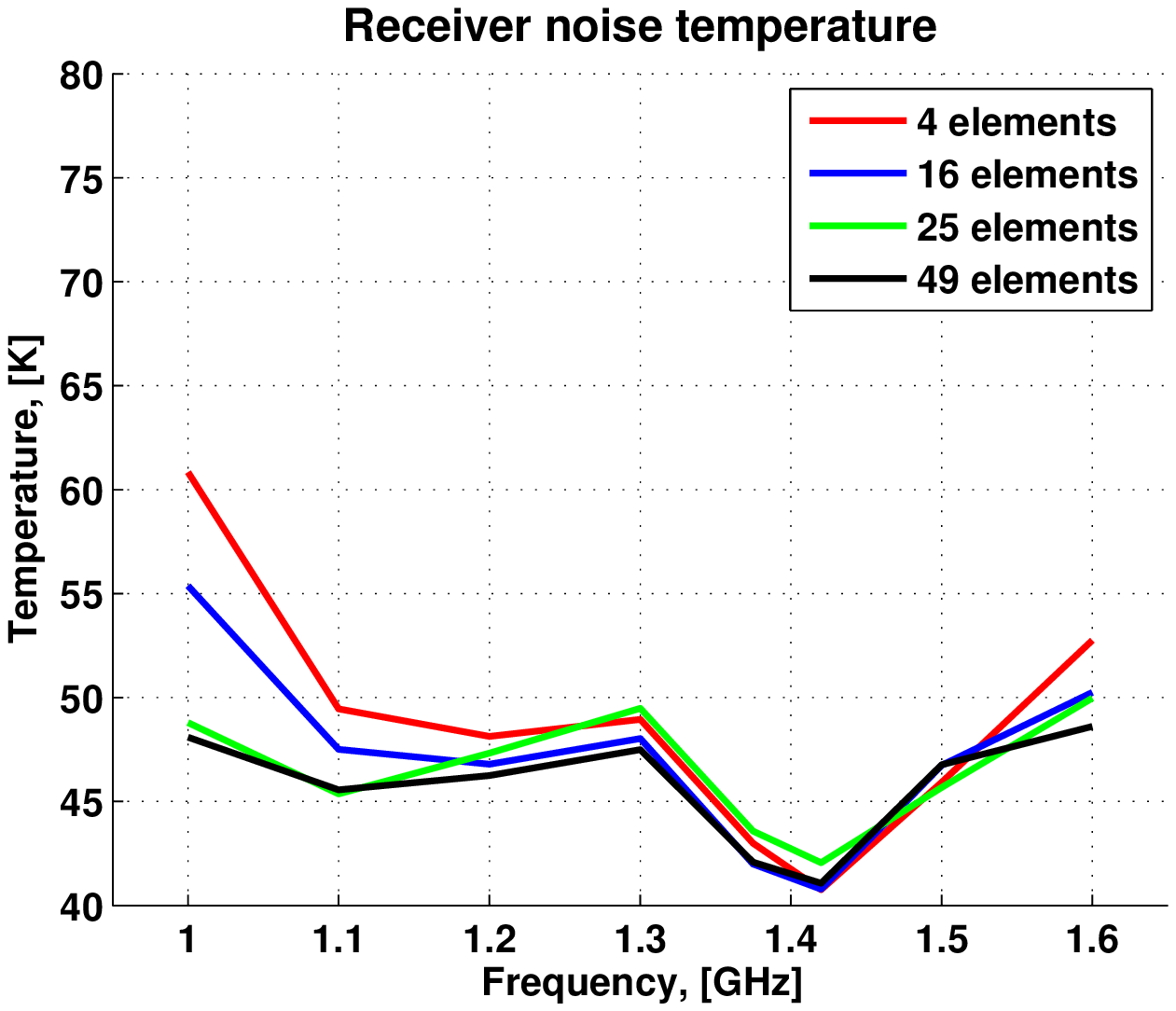} & \includegraphics[width=8cm]{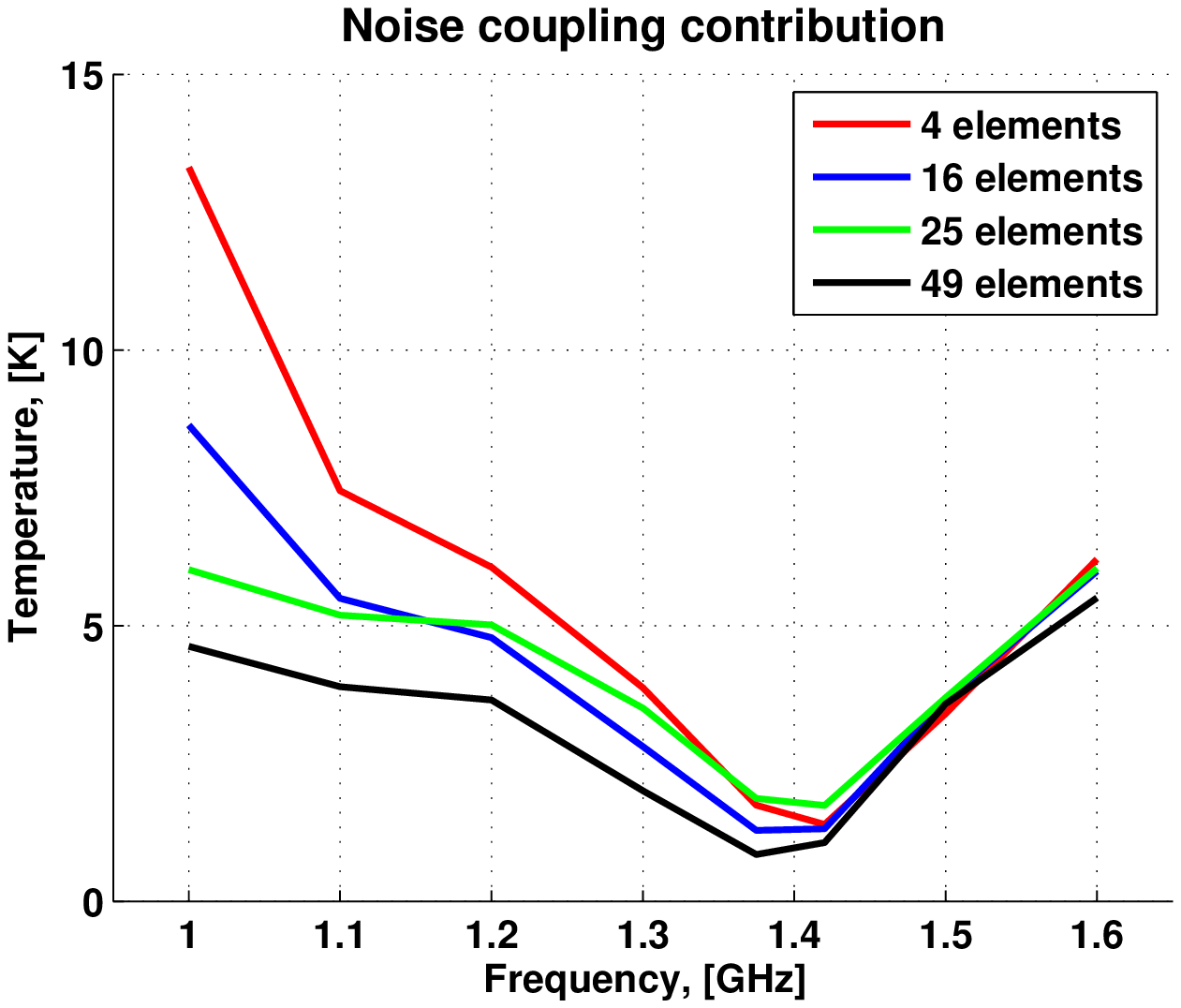} \\
\small (a)  & \small (b)  \\\\
\includegraphics[width=8 cm]{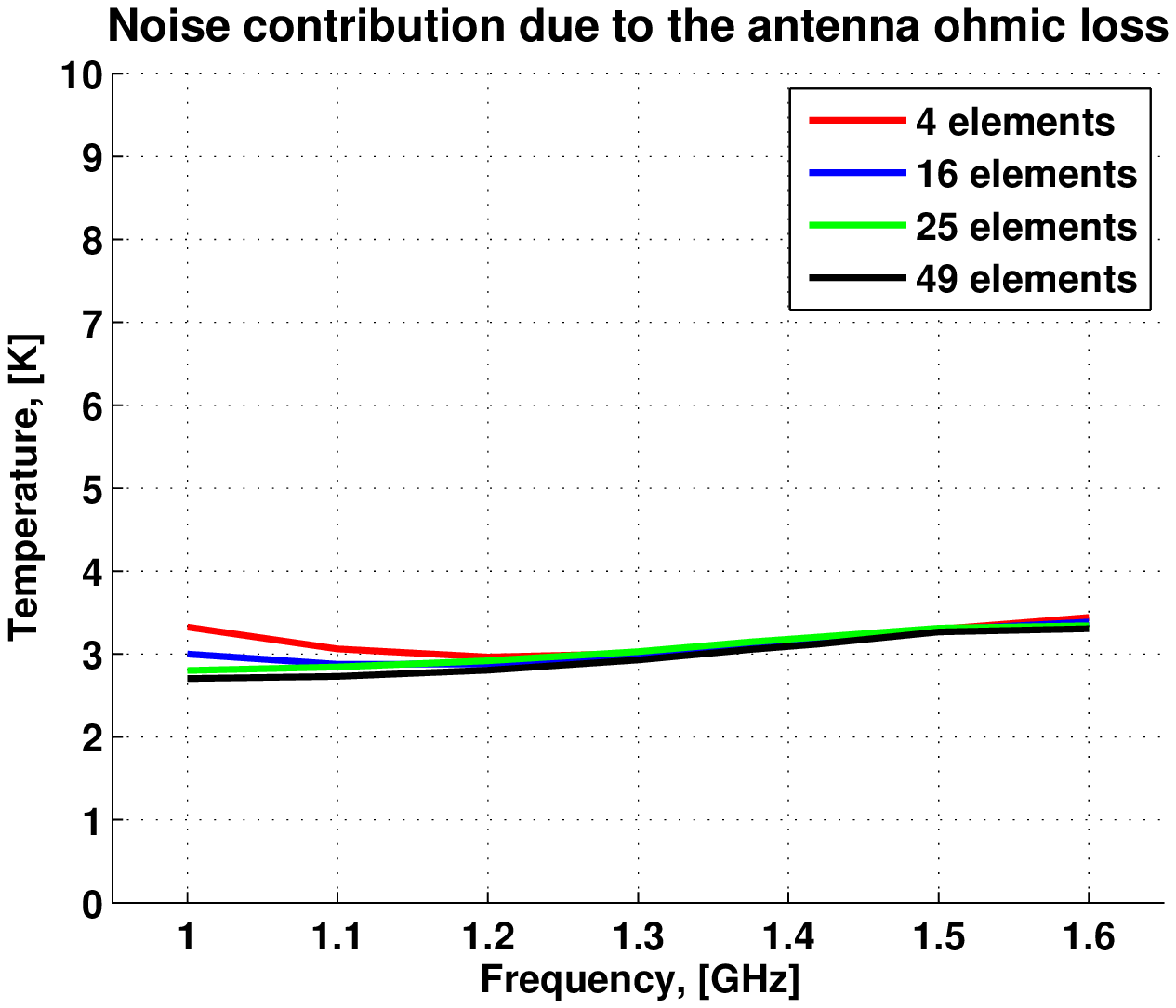} & \includegraphics[width=8cm]{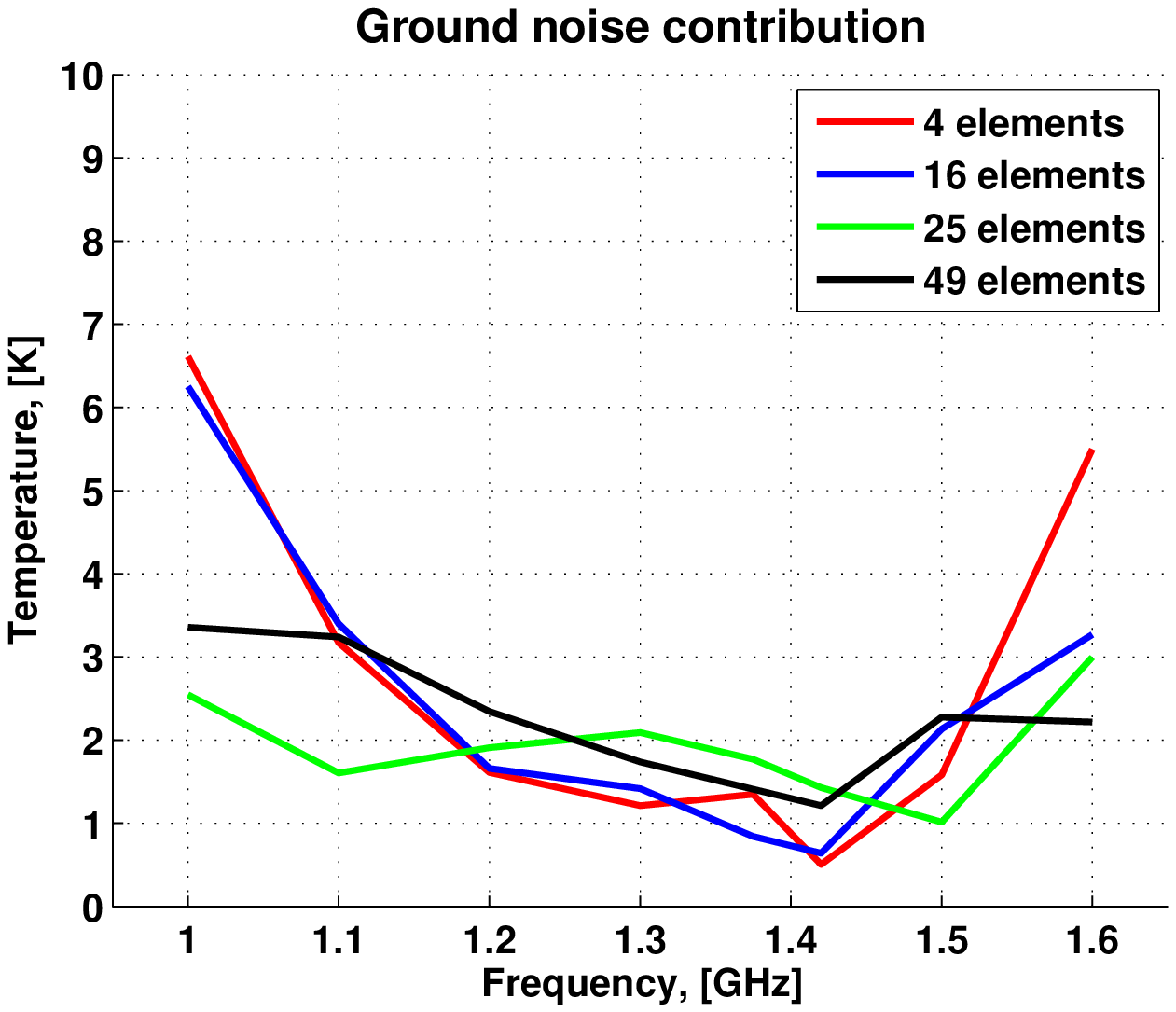} \\
\small (c)  & \small (d)  \\
\end{tabular}
\caption{(a) The simulated receiver noise temperature versus
frequency and its contributions due to (b) the noise coupling
effects in the receiver, (c) the ohmic losses of the antenna and (d)
the external noise due to the ground (due to the back radiation of
the array, when $\theta\geq 90^{\text{o}}$). All temperatures are
for broad side beams.} \label{ExperimArray_Trec_vs_freq}
\end{figure}

In this section, we present the numerical results
for practical beamformers for the frequencies ranging from 1 GHz to
1.6 GHz and over the scan range within which $\theta$ changes from
$0^{o}$ to $85^{o}$ and $\phi=0-360^{o}$. The practical beamformers
combine the signals received by 4, 16, 25 and 49 elements. The numerical results
include the directivity of the array
antenna (see Fig.~\ref{ExpArray_weights} and the receiver noise temperature and
its contributions due to noise coupling effect, antenna ohmic loss,
and ground noise due to back radiation (see
Fig.~\ref{ExperimArray_Trec_vs_freq}--\ref{ExperimArray_Trec_vs_3Dscan_1p6GHz}).

\begin{figure}[!ht]
\centering
\includegraphics[width=8 cm]{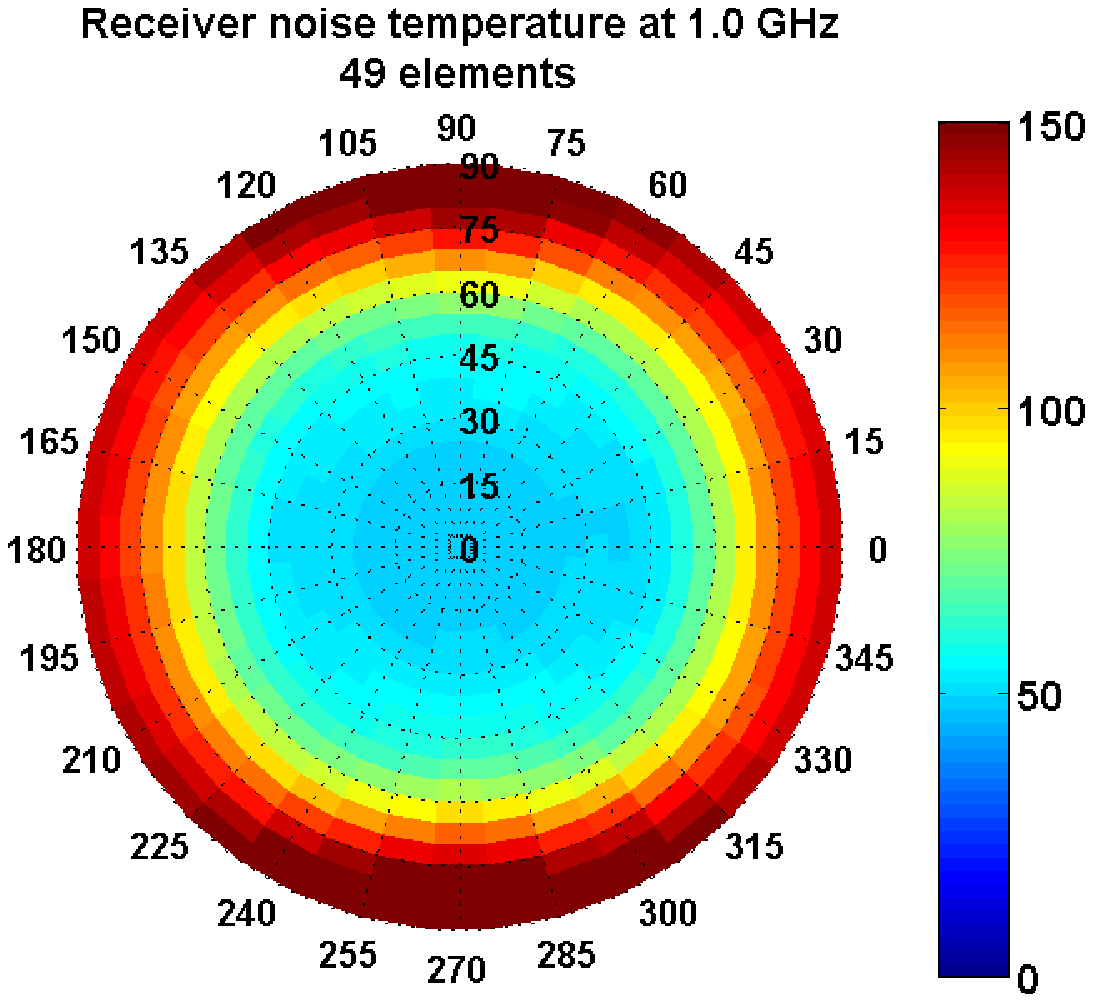}\\
\small (a)    \\
\begin{tabular}{cc}
\includegraphics[width=8 cm]{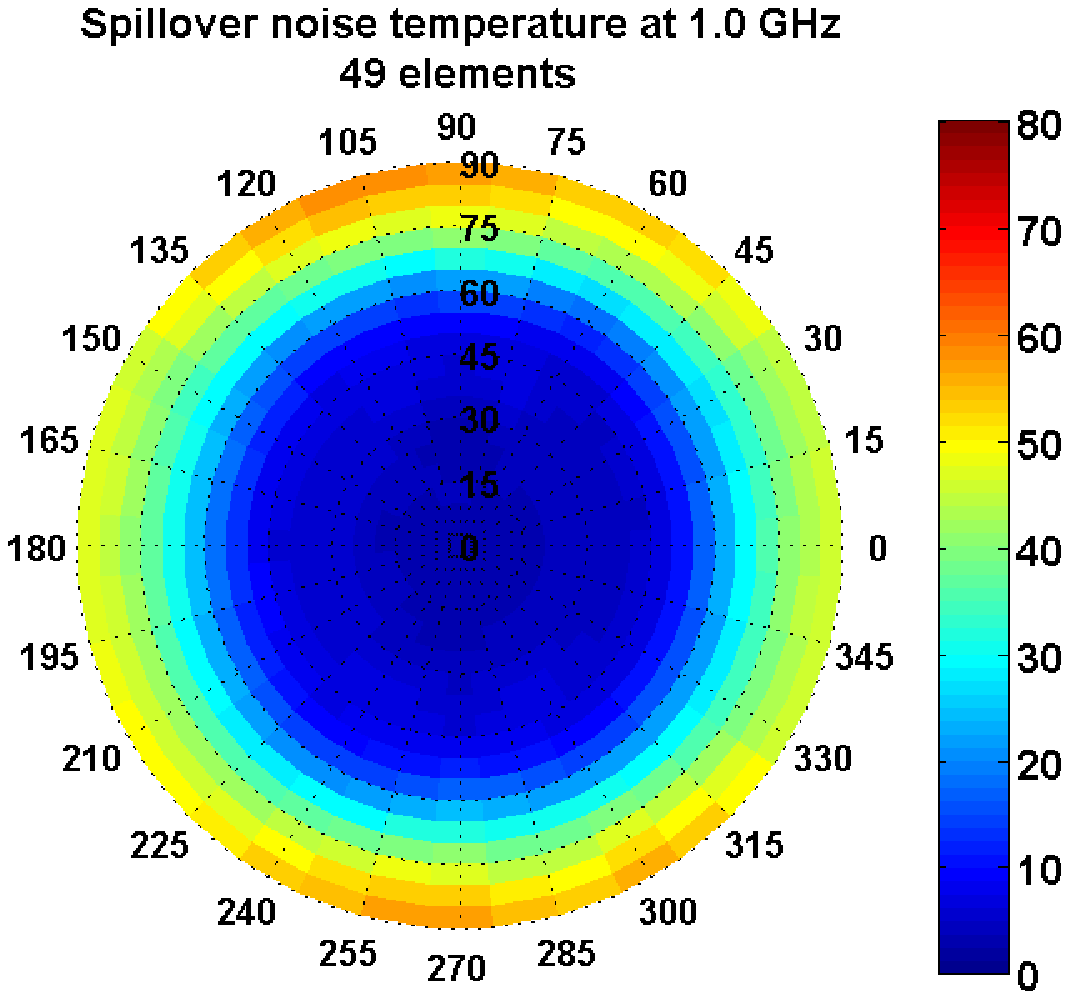} & \includegraphics[width=8cm]{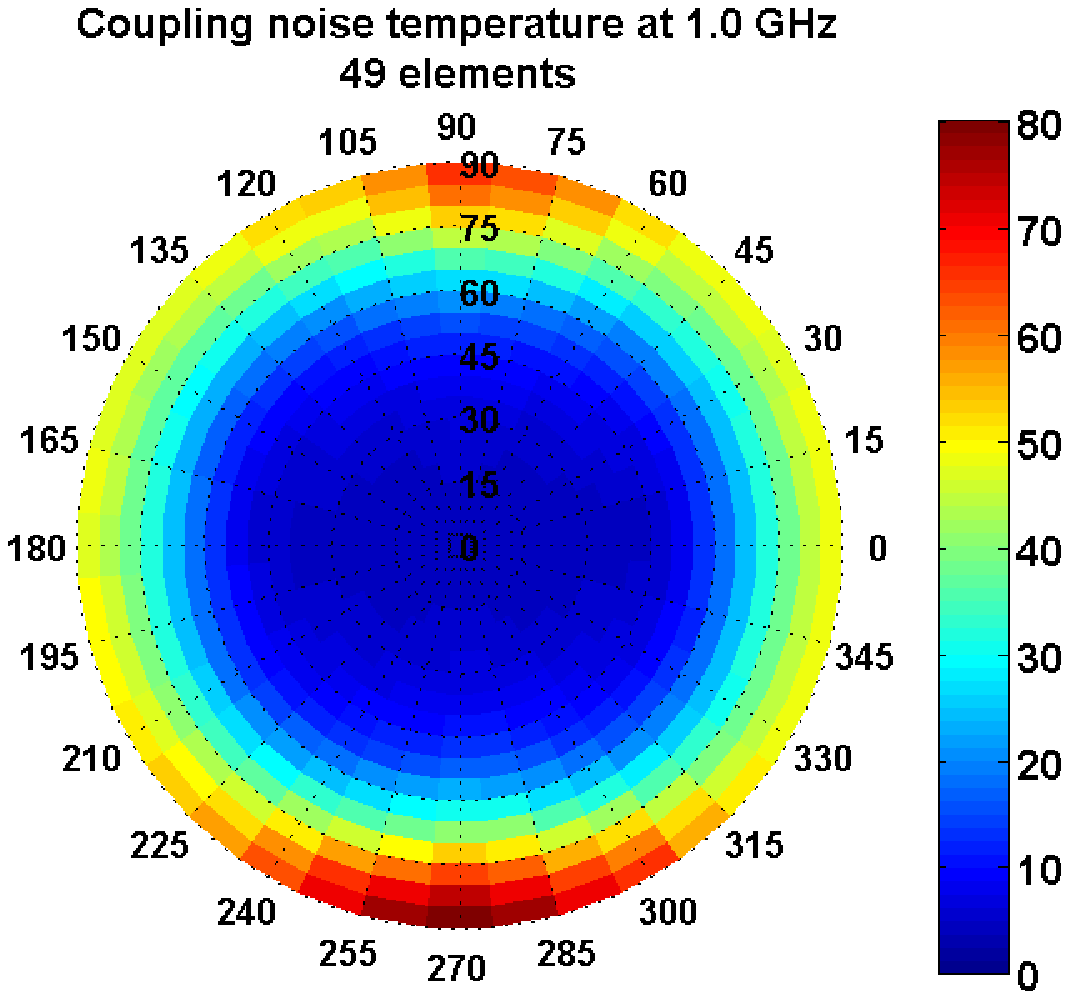}\\
\small (b) & \small (c)  \\
\end{tabular}
\caption{(a) The simulated receiver noise temperature ([K]) over 
scan angle at 1.0 GHz and its contributions due to (b) the noise
coupling effects in the receiver, (c) the ohmic losses of the
antenna and (d) the external noise due to the ground (due to the
back radiation of the array, when $\theta\geq 90^{\text{o}}$).}
\label{ExperimArray_Trec_vs_3Dscan_1p0GHz}
\end{figure}

Figure~\ref{ExperimArray_Trec_vs_freq}(a) shows the receiver noise
temperatures of the DIGESTIF tile that were computed for the
boresight direction of observation at 8 frequency points within the
bandwidth of 1-1.6 GHz. These results clearly demonstrate that for
all practical beamformers the on-axis beam noise temperature is
weakly dependent on frequency and takes values between 42 and 61 K
that are 5-30\% higher than the minimum noise temperature of LNAs
($T_\text{min}=35-40$)K. The noise contributions due to the receiver
noise coupling effects $T_\text{coup}$, antenna ohmic losses
$T_\text{rad}$ and external (ground) noise pick-up $T_\text{ext}$,
as shown on fig.~\ref{ExperimArray_Trec_vs_freq}(b),
~\ref{ExperimArray_Trec_vs_freq}(c) and
~\ref{ExperimArray_Trec_vs_freq}(d) do not exceed 13, 3.5 and 6.5 K,
respectively. At 1.42 GHz - the frequency at which the array design
was optimized - the temperatures $T_\text{coup}$ and $T_{ext}$ and
the total receiver temperature take the minimum values within the
operational bandwidth as the result of the minimized impedance
mismatch loss and relatively low side and back radiation levels.
\begin{figure}[!ht]
\centering
\begin{tabular}{cc}
\includegraphics[width=8 cm]{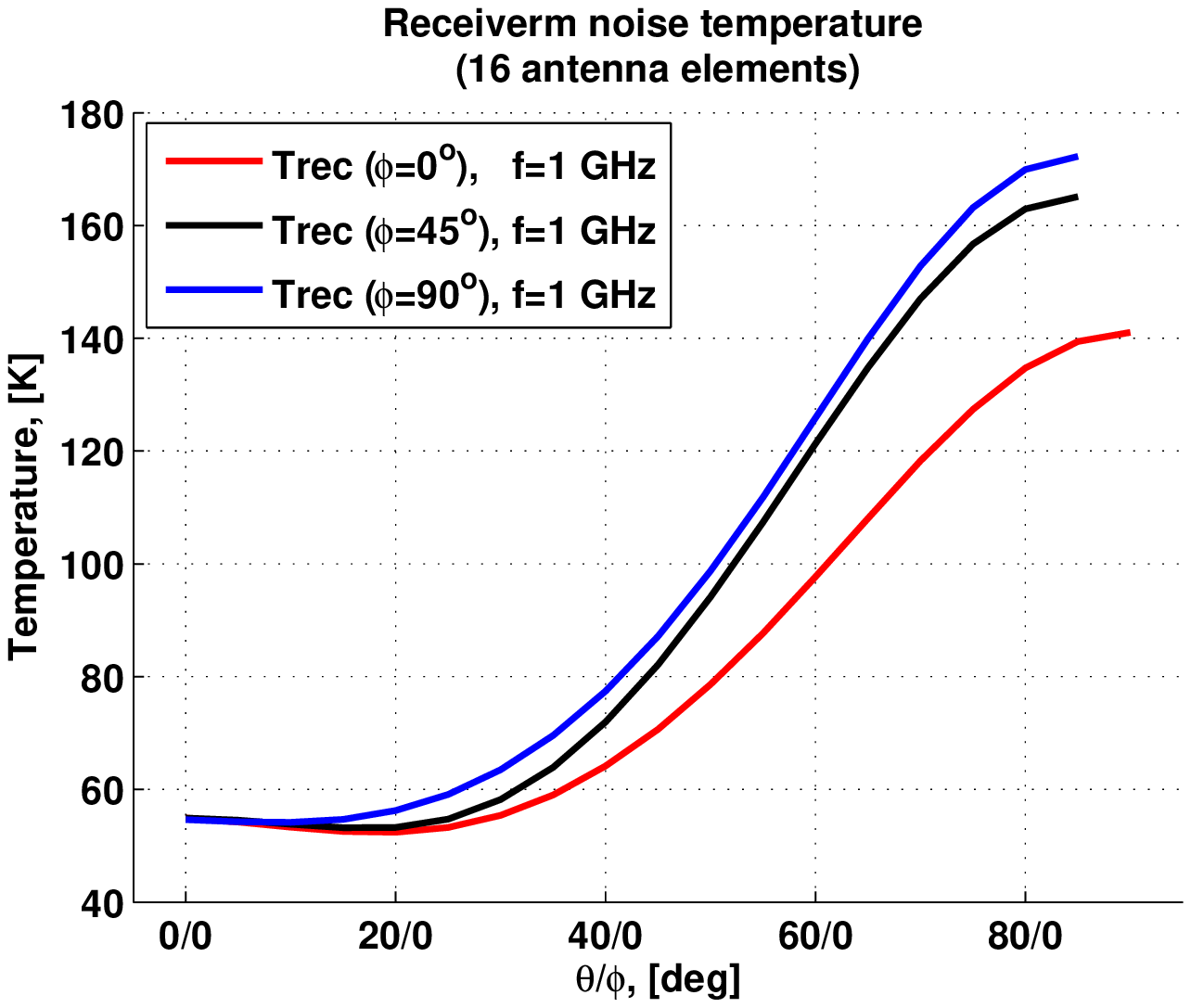} & \includegraphics[width=8cm]{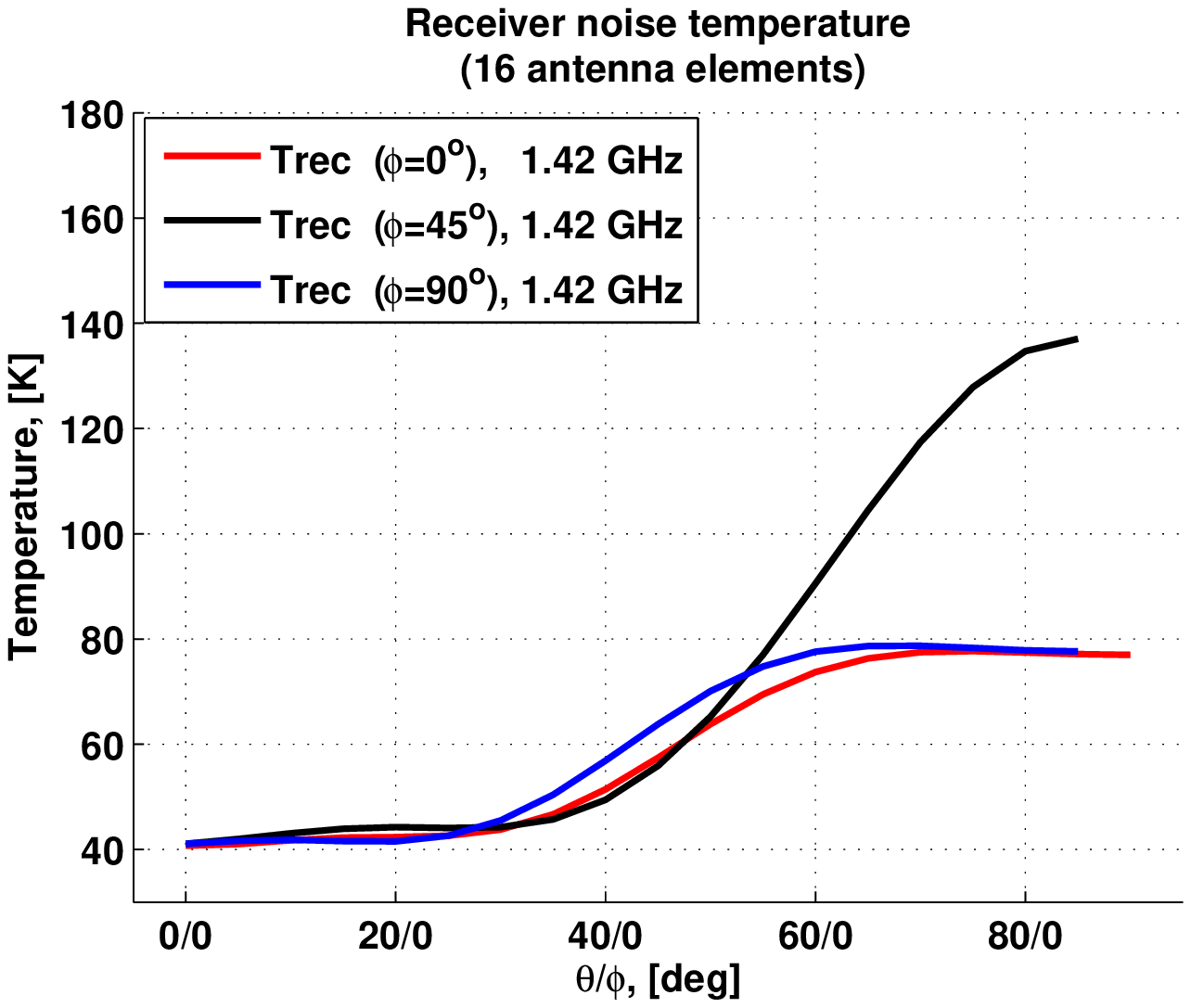} \\
\small (a)  & \small (b)  \\\\
\includegraphics[width=8 cm]{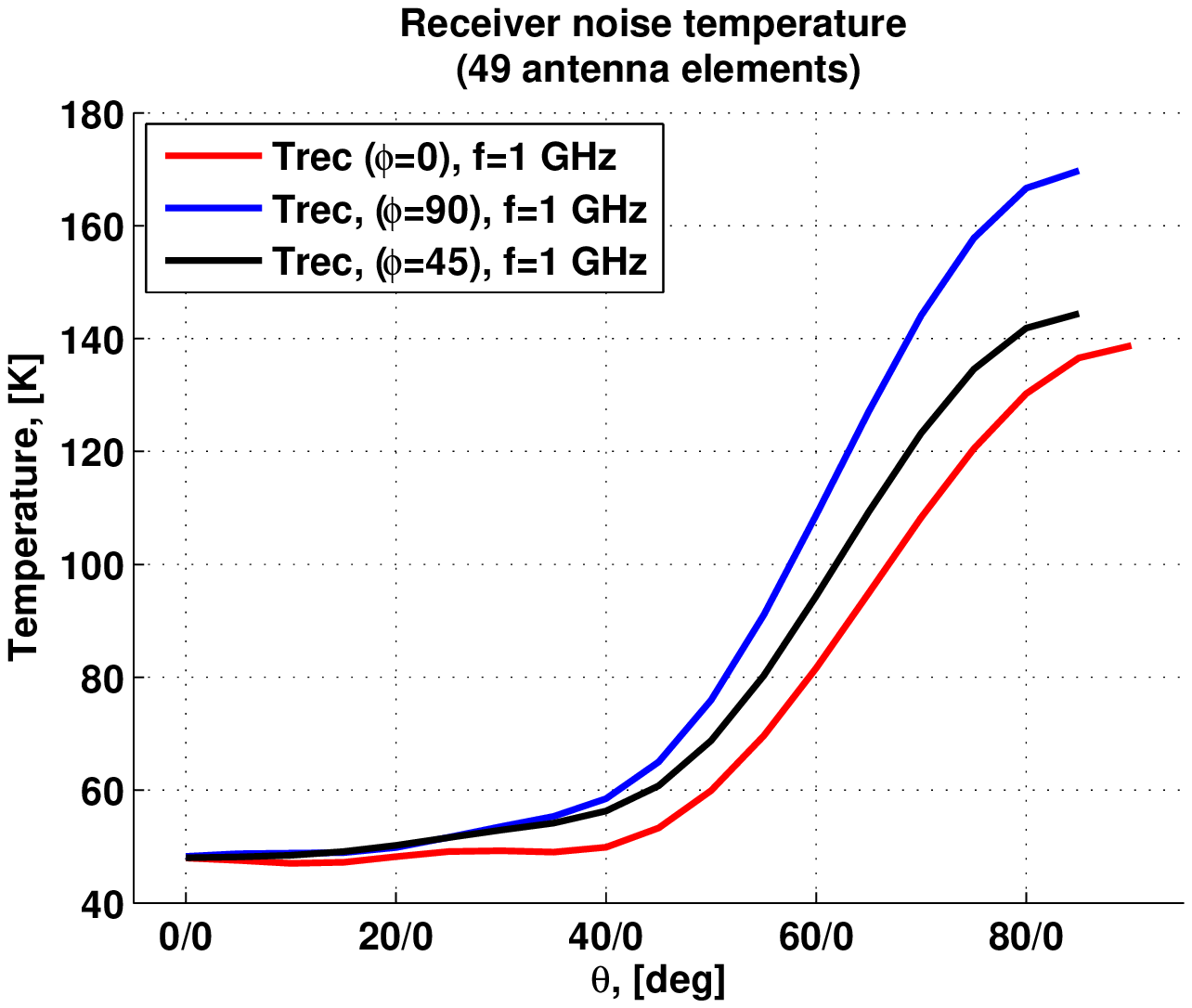} & \includegraphics[width=8cm]{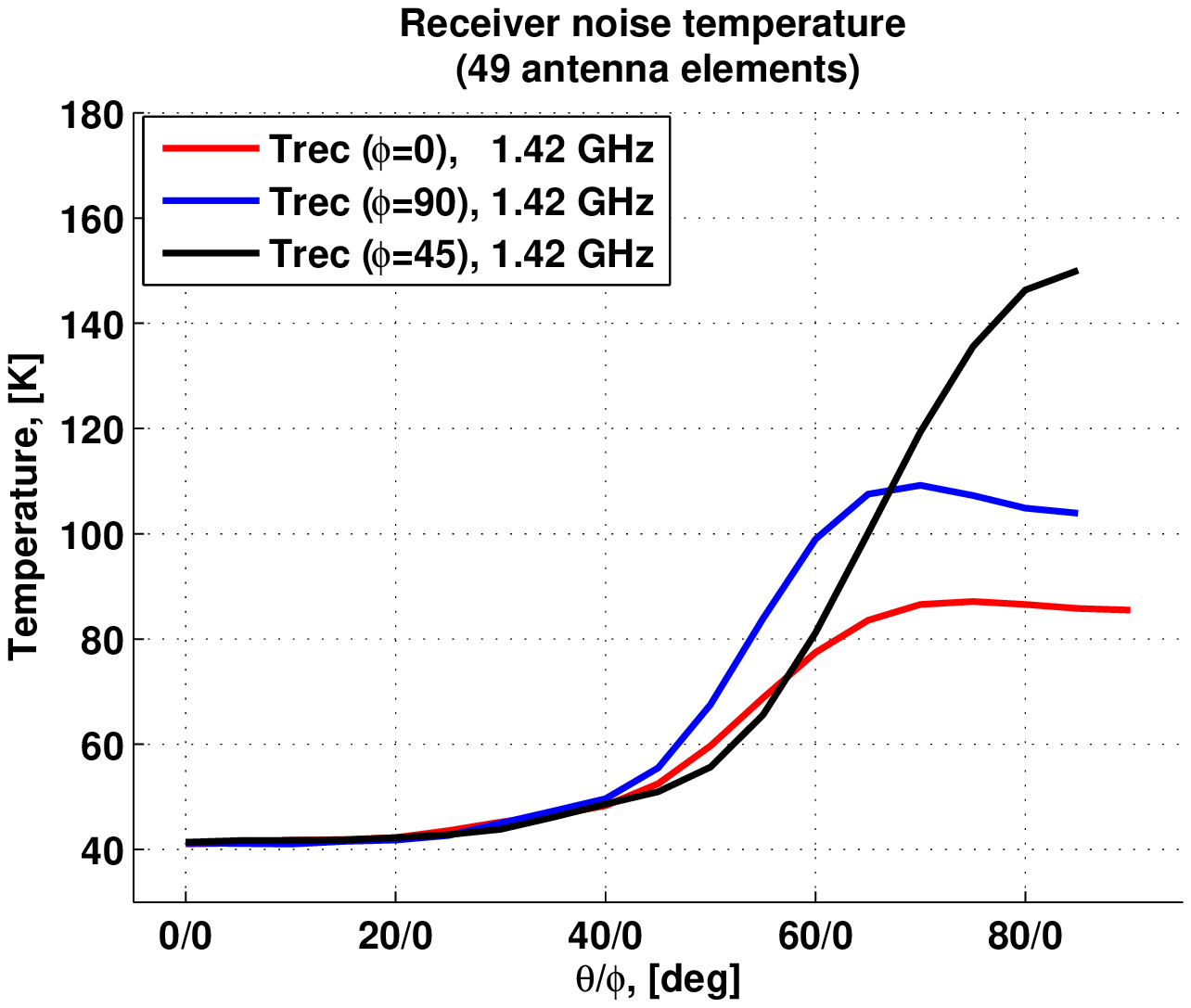} \\
\small (c)  & \small (d)  \\
\end{tabular}
\caption{The simulated receiver noise temperatures versus scan angle
in three scan planes for beamformers with (a,b) 16 and (c,d) 49
active antenna elements at 1.0 GHz and 1.42 GHz.}
\label{Trec_3ScanCuts_1p0GHz_16el}
\end{figure}
Figure~\ref{Trec_3ScanCuts_1p0GHz_16el}(a)-(d) shows how the
receiver noise and its weight-dependent noise components vary with
scan angle for beamformers with 16 and 49 elements at 1 GHz and 1.42
GHz. For these beamformers, respectively, the increase of the noise
temperatures is less than 20\% when the scan angle is smaller than
$\sim30^\text{o}$ and $\sim40^\text{o}$ off boresight direction. For
larger scan angles, however, $T_\text{rec}$ rapidly increases and
becomes as high as 80-160 K depending on the number of active
antenna elements, scan plane and frequency. Such high values are
mainly due to the strong mutual coupling between antenna elements at
low frequencies (causing the rise of the receiver noise coupling
contribution as observed on
fig.~\ref{ExperimArray_Trec_vs_3Dscan_1p0GHz}(c)) and high side-lobe
levels at high frequencies for scanned beams.



\subsection{Comparison with measurements in the open environment at Westerbork}

This subsection compares the predicted and measured receiver
noise temperatures versus frequency. The
results of cross-comparison of the measured Trec at Westerbork (in
the open environment) and inside THACO (which is expected to shield
the receiver from the ground noise) are presented. Also, the
measured and modeled noise temperatures of a single Vivaldi element
receiver inside THACO are shown.

\begin{figure}[!ht]
\centering
\begin{tabular}{cc}
\includegraphics[width=8 cm]{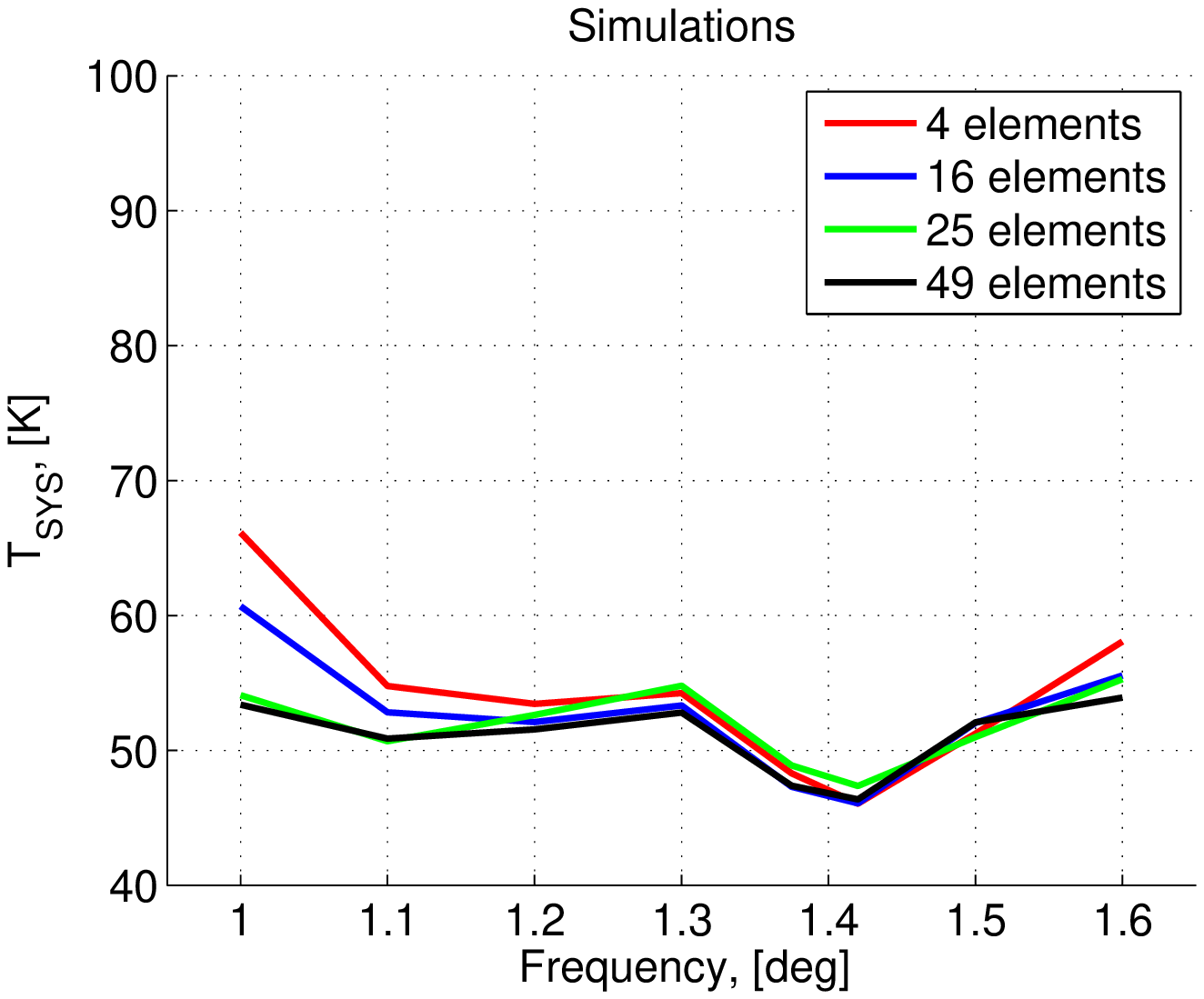} & \includegraphics[width=8cm]{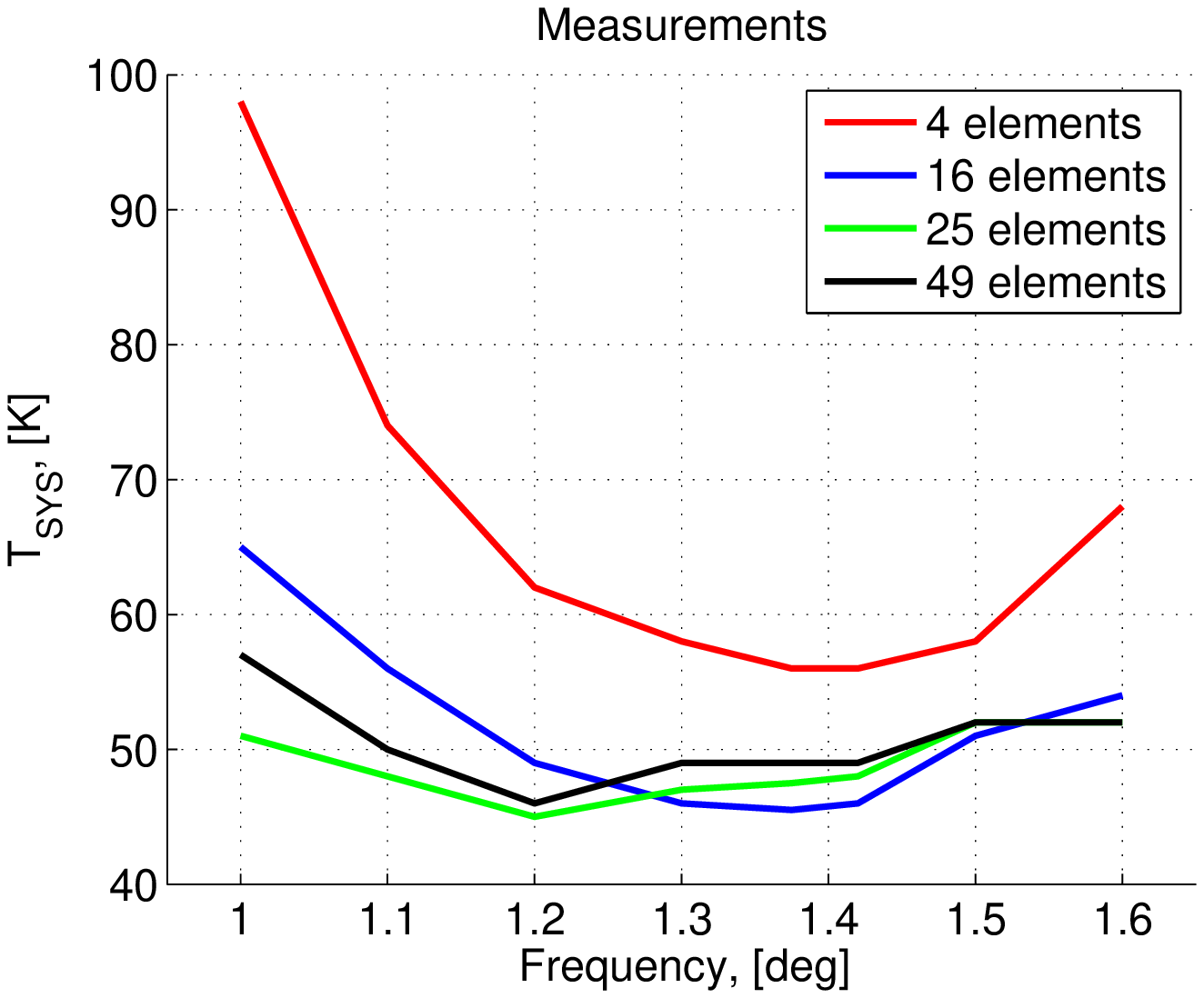}
\\\\
\small (a)  & \small (b)  \\
\end{tabular}
\includegraphics[width=8 cm]{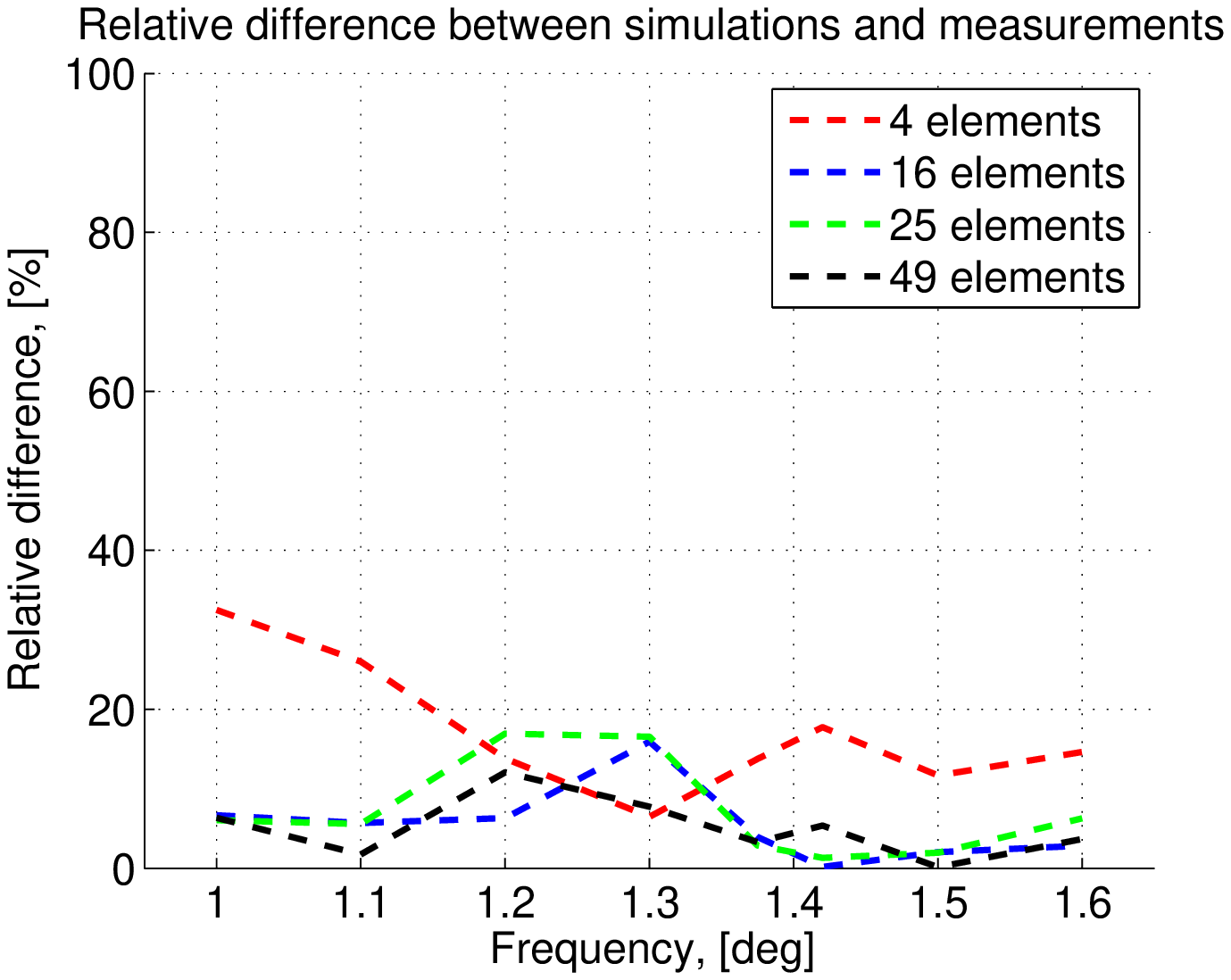} \\
\small (c) \\
\caption{(a) The simulated and (b) measured system noise
temperatures ($T_\text{SYS}=T_\text{rec}+T_\text{sky}$) versus
frequency and (c) the relative difference between them.}
\label{ExperimArray_SIM_Trec_vs_freq}
\end{figure}

\begin{figure}[!ht]
\centering
\begin{tabular}{cc}
\includegraphics[width=8 cm]{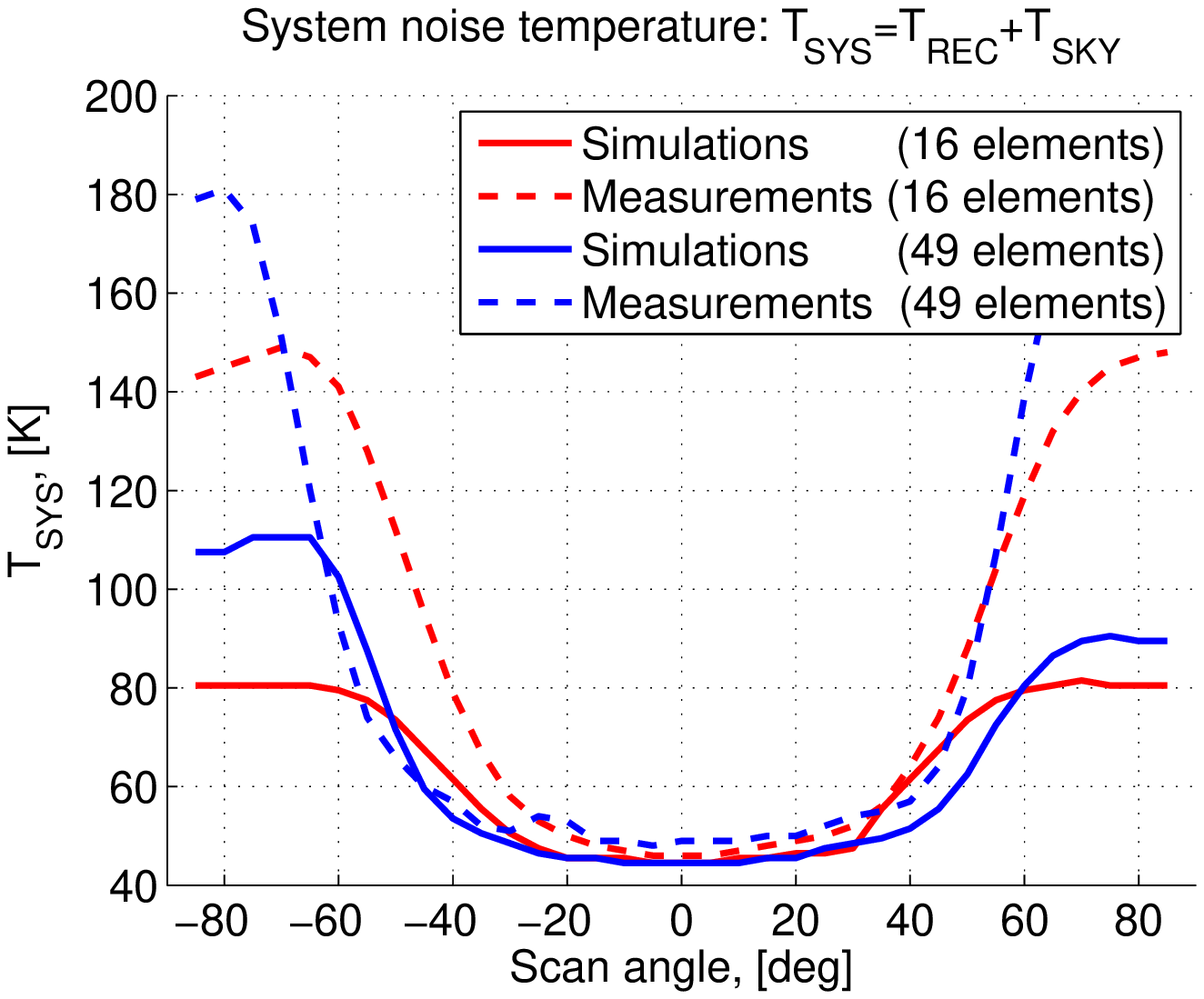} & \includegraphics[width=8cm]{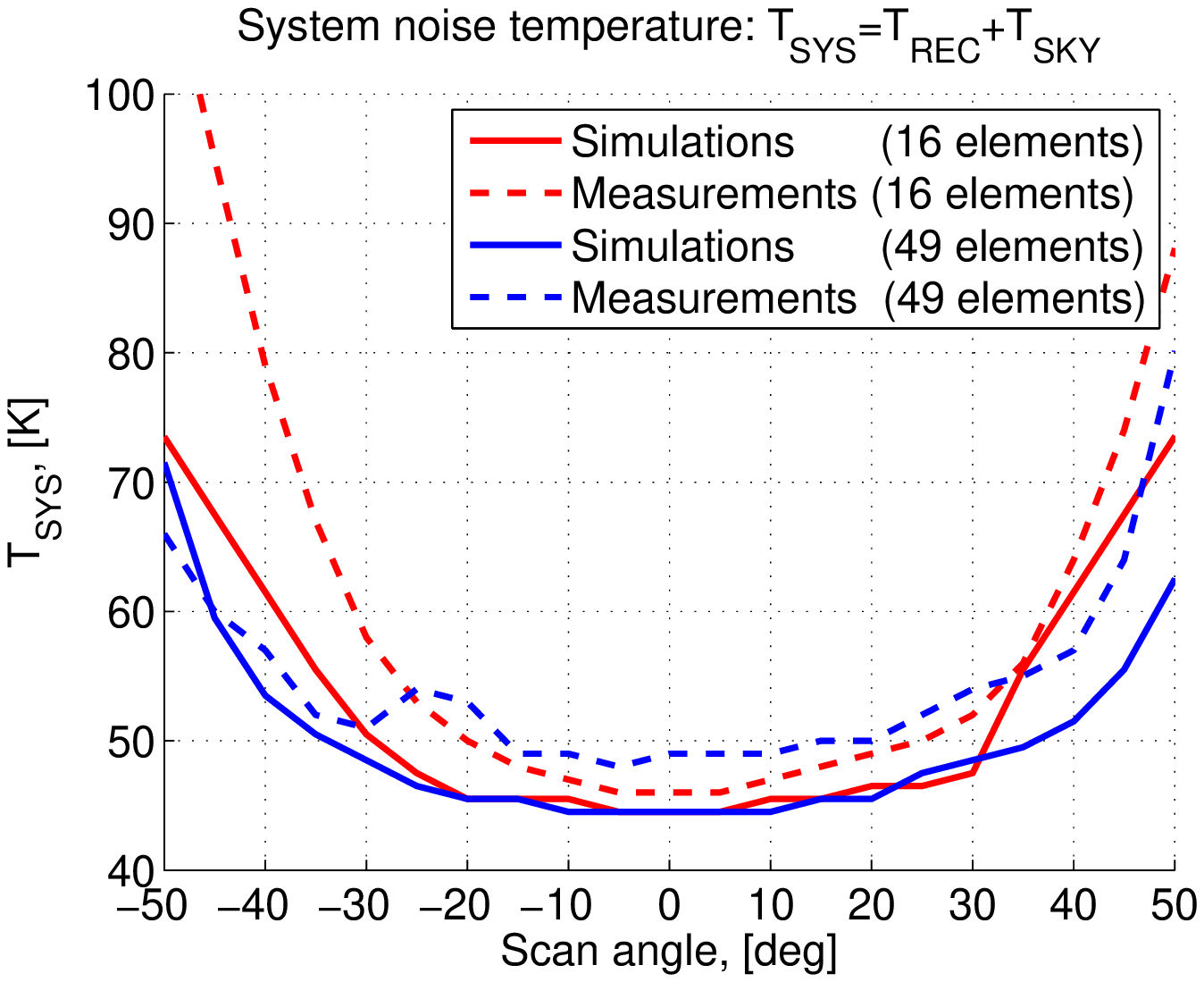}
\\\\
\small (a)  & \small (b)  \\
\end{tabular}
\includegraphics[width=8 cm]{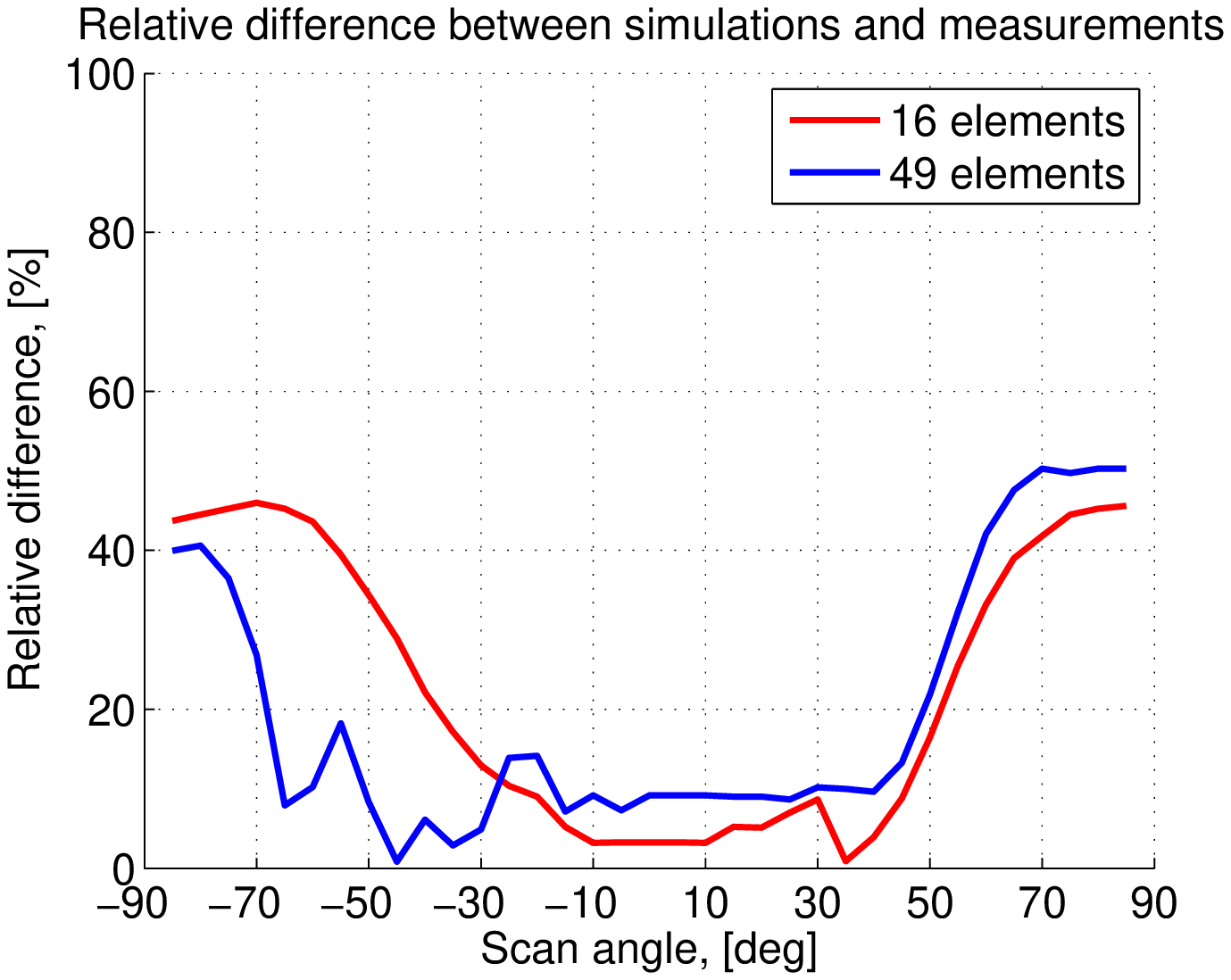} \\
\small (c) \\
\caption{The simulated and measured receiver noise temperatures
versus scan angle for (a) a large and (b) small scan range and (c)
the relative difference between them.}
\label{ExperimArray_Large_Trec_vs_ScanAngle}
\end{figure}

\begin{figure}[!ht]
\centering
\begin{tabular}{cc}
\includegraphics[width=8 cm]{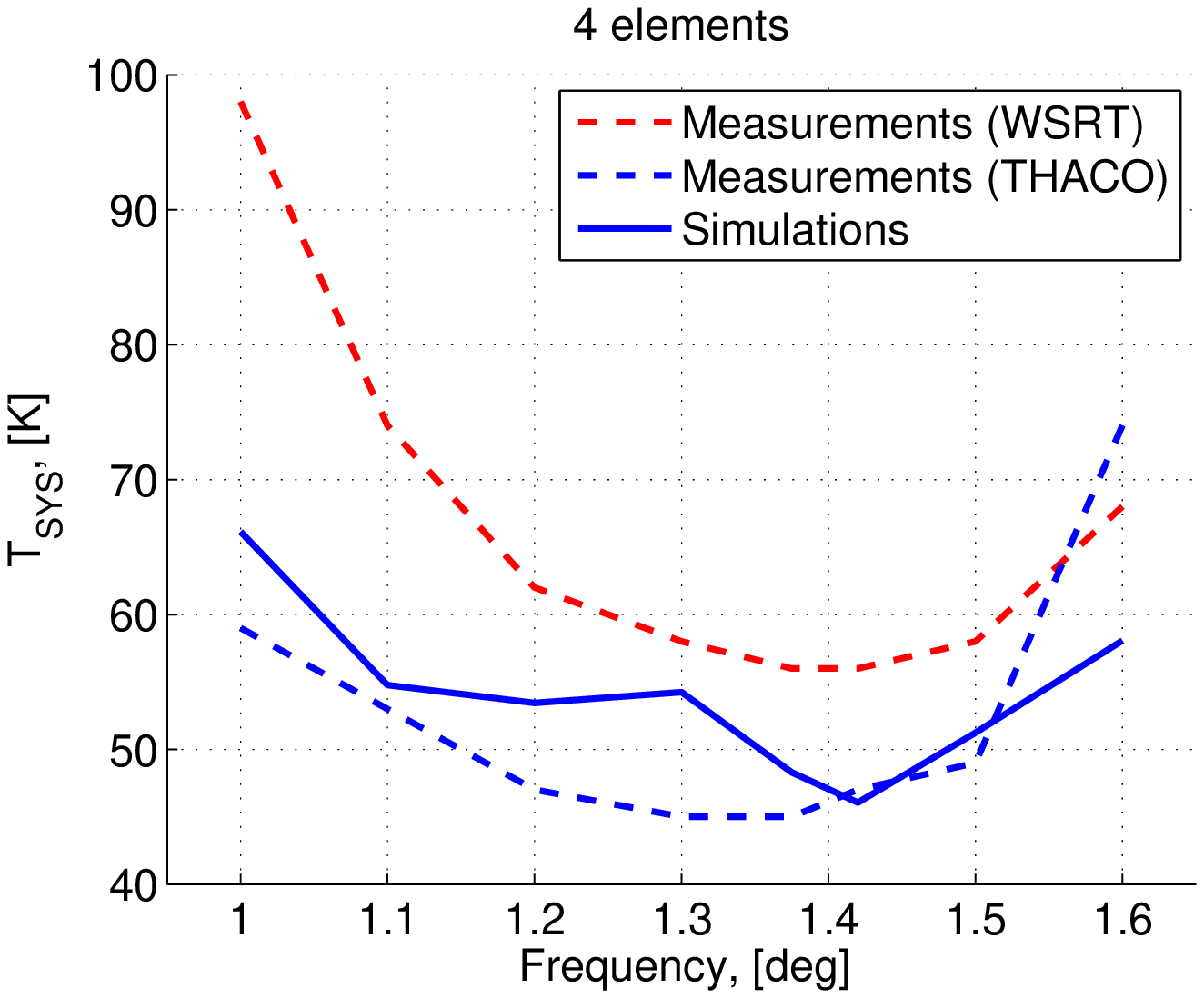} & \includegraphics[width=8cm]{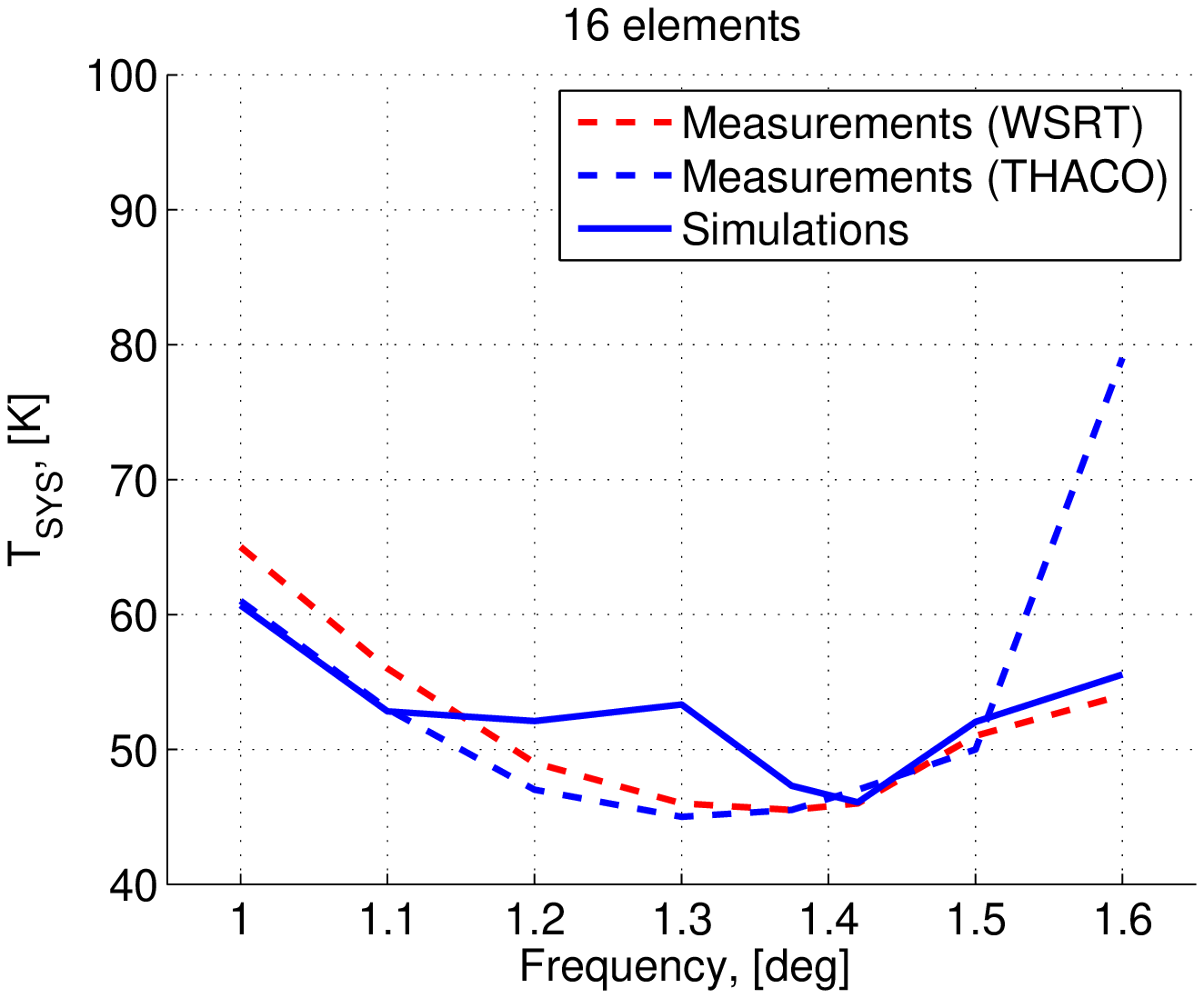} \\
\small (a)  & \small (b)  \\\\
\includegraphics[width=8 cm]{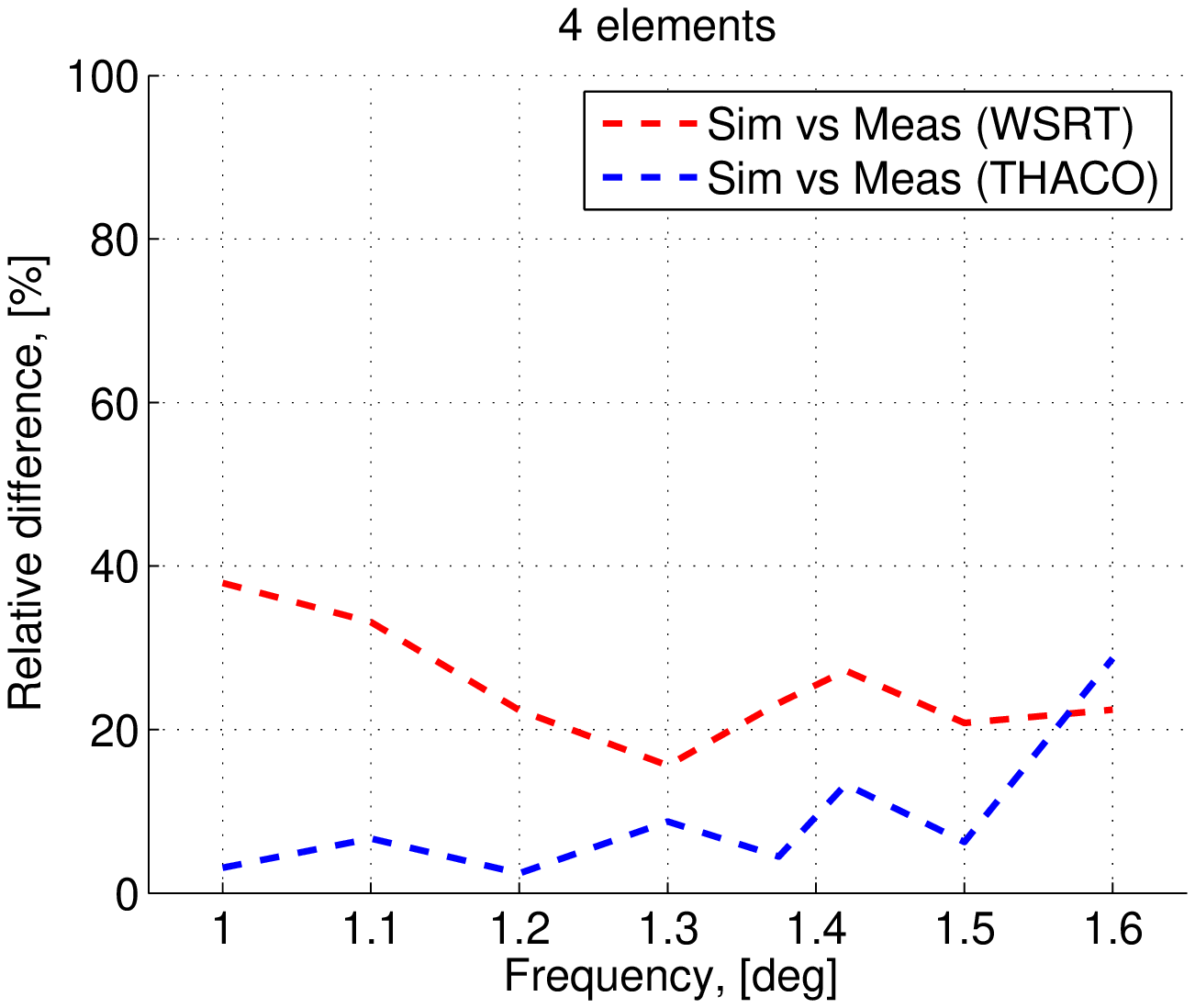} & \includegraphics[width=8cm]{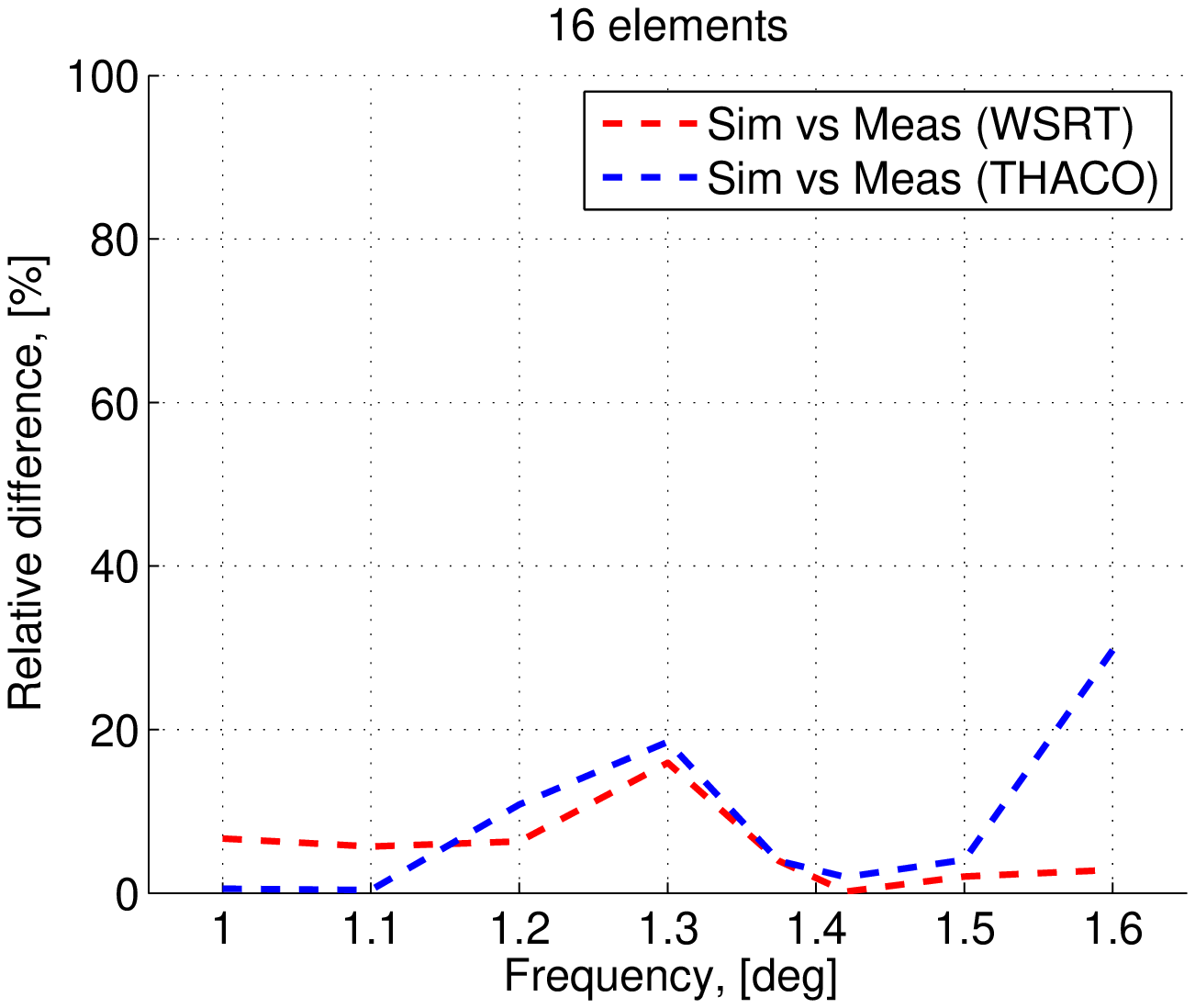} \\
\small (c)  & \small (d)  \\
\end{tabular}
\caption{The simulated and measured receiver noise temperatures
versus frequency for (a) 4-element and (b) 16-element beamformers
and (c,d) the relative difference between the measurements and
simulations for the corresponding beamformers. }
\label{THACO_Rrec_vs_freq}
\end{figure}

Upon collecting the simulation and measurement results, we can
compute the relative difference between the modeled and measured
receiver noise temperatures.
Figures~\ref{ExperimArray_SIM_Trec_vs_freq}-\ref{ExperimArray_Large_Trec_vs_ScanAngle}
show both the simulated and measured $T_{rec}$ as well as their
relative difference as a function of frequency and scan angle. As
one can see on fig.\ref{ExperimArray_SIM_Trec_vs_freq}(c), the
relative difference between simulations and measurements is smaller
than 20\% over the entire frequency range for all practical
beamformers, except for the beamformer with 4 active channels in the
region of 1-1.2 GHz. At these frequencies the 4-element subarray has
a rather low gain ($<$10 dB) and, thus the experimental receiver
picks up the noise due to the buildings and trees that are present
in the actual environment, but were not accounted for in the model.
This reasoning is supported by the measurements which were done
inside the shielding cabin (THACO) (see
fig.\ref{THACO_Rrec_vs_freq}(c)). For the latter tests, the measured
noise temperatures for the 4-channel beamformer lie within the
region of the simulated values with 15-20\% difference at most
frequencies.

The agreement between the simulated and measured noise temperatures
over the scan range is also good, and for the beamformers with 16
and 49 channels was found to be at the level of $<$25\% within the
scan range of $\pm (40-45^\text{o})$. For larger off-axis angles,
the measured temperatures are much higher with the difference up to
50\% relatively to the corresponding simulated values.

\newpage
\subsection{Comparison with measurements inside THACO}

\newpage
\newpage
\section{Conclusions}
The noise performance of the aperture array tile receiver has been
simulated and compared to the experimental results as obtained
though the `hot-cold' measurement procedure inside a shielding cabin
(THACO) and in the open environment near WSRT. The measurements have
been carried out for several practical beamformers (with 4, 16, 25
and 49 active channels) over the frequency band from 1 to 1.6 GHz
and a wide 3D beam scan range. The presented numerical results
include the antenna patterns and system noise temperatures $T_{sys}$
for all considered practical situations, as well as separate noise
contributions due to the receiver noise coupling effects
$T_\text{coup}$, antenna ohmic loss $T_\text{rad}$ and external (sky
and ground) noise $T_\text{ext}$.

The numerical results demonstrate that the on-axis beam noise
temperatures take values ranging between 42 and 61 K within the
frequency bandwidth and are maximum 30\% higher than the minimum
noise temperature of LNAs ($T_\text{min}=35-40$)K. The noise
contributions $T_\text{coup}$, $T_\text{rad}$ and $T_\text{ext}$ do
not exceed 13, 3.5 and 6.5 K, respectively for broad side beams. At 1.42 GHz - the
frequency at which the antenna was optimized - $T_{sys}$ is lowest,
as the result of the minimized impedance mismatch loss and
relatively low side and back radiation levels.

The receiver noise temperature exhibits a strong dependence on the
beamformer weights and degrades when scanning far off boresight
direction. For beamformers with 16 and 49 channels respectively, the
relative increase of $T_{rec}$ was found to be less than 20\% for
the scan angles smaller than $\sim30^\text{o}$ and
$\sim40^\text{o}$, and a factor 2-4 for larger angles, depending on
frequency and beamformer. Such high values of $T_{rec}$ are mainly
due to the strong mutual coupling between antenna elements at low
frequencies (causing the rise of $T_\text{coup}$ and high side-lobe
levels at high frequencies for scanned beams.

There is a good agreement between simulations and measurements that
were performed in the open environment at Westerbork. The relative
difference between the modeled and measured $T_\text{SYS}$ is
smaller than 20-25\% over the entire frequency band and within the
scan range of $\pm (40-45^\text{o})$, except for the 4-channel
beamformer for which this difference can be twice as large. For the
latter beamformer case, the antenna pattern is rather broad (the
gain is $<$10 dB) and, thus the experimental receiver can pick up an
additional noise component due to the buildings and trees that are
present in the actual environment, but were not accounted for in the
model. This reasoning is supported by the measurements inside the
shielding cabin (THACO) for which the agreement with simulations
significantly improves. Furthermore, for large off boresight scan
angles, the temperatures as measured in the open environment are much higher
than the simulated ones, most likely due to the above mentioned
effects of the noisy environment of which the exact temperature
distribution is not well known.

\section{Acknowledgements}
This work has been supported in part by  the Netherlands Organization for Scientific Research (APERTIF project funded by NWOGroot),
the Swedish Agency for Innovation Systems VINNOVA and Chalmers University of Technology (VINNMER - Marie Curie Actions international qualification fellowship).

\newpage
\bibliographystyle{IEEEtran}
\bibliography{IEEEabrv,mybibfile_old}

\end{document}